\documentclass[prl,superscriptaddress,showpacs,longbibliography,reprint]{revtex4-2} 

\usepackage{hyperref}
\usepackage{color}
\usepackage[utf8]{inputenc}
\usepackage[usenames,dvipsnames]{xcolor}
\usepackage{amsmath}
\usepackage{amssymb}
\usepackage{graphicx}
\graphicspath{{graphics/}}
\usepackage{epsfig}
\usepackage{dcolumn}
\usepackage{bm}
\usepackage{mathrsfs}
\usepackage{multirow}
\usepackage[all]{xy}
\usepackage{pbox}
\usepackage{lipsum}
\usepackage{verbatim}
\usepackage{braket}
\usepackage{dsfont}
\usepackage{array}
\usepackage{makecell}
\usepackage{tabularx}
\usepackage{isotope}
\usepackage{xr}

\DeclareMathOperator{\Tr}{Tr}

\usepackage{graphicx}
\graphicspath{{./Figures/}}

\usepackage{color,soul}

\begin{document}
\title{Exact solution  of a boundary time-crystal phase transition: time-translation symmetry breaking and non-Markovian dynamics of correlations}
\author{Federico Carollo}
\affiliation{Institut f\"ur Theoretische Physik, Universit\"at Tübingen, Auf der Morgenstelle 14, 72076 T\"ubingen, Germany}
\author{Igor Lesanovsky}
\affiliation{Institut f\"ur Theoretische Physik, Universit\"at Tübingen, Auf der Morgenstelle 14, 72076 T\"ubingen, Germany}
\affiliation{School of Physics and Astronomy and Centre for the Mathematics and Theoretical Physics of Quantum Non-Equilibrium Systems, The University of Nottingham, Nottingham, NG7 2RD, United Kingdom}

\begin{abstract}
The breaking of the continuous time-translation symmetry manifests, in Markovian open quantum systems, through the emergence of non-stationary dynamical phases. Systems that display nonequilibrium transitions into these phases are referred to as time-crystals, and they can be realized, for example, in many-body systems governed by collective dissipation and long-ranged interactions. Here, we provide a complete analytical characterization of a boundary time-crystal phase transition. This involves exact expressions for the order parameter and for the dynamics of quantum fluctuations, which, in the time-crystalline phase,  remains asymptotically non-Markovian as a consequence of the time-translation symmetry breaking. We demonstrate that boundary time-crystals are intrinsically critical phases, where fluctuations exhibit a power-law divergence with time. Our results show that a dissipative time-crystal phase is far more than merely a classical non-linear and non-stationary (limit cycle) dynamics of a macroscopic order parameter. It is rather a genuine many-body phase where the properties of correlations distinctly differs from that of stationary ones. 
\end{abstract}

\maketitle

\noindent {\bf Introduction.---} Dissipation and irreversible effects are typically associated with the convergence of a quantum system towards an asymptotic stationary state. Effectively time-translation invariant states can also occur in closed quantum systems when considering the dynamics of local observables \cite{rigol2008,polkovnikov2011,dalessio2016}. Recently, this paradigm has been challenged by the observation that non-stationary asymptotic behavior can emerge in open quantum systems \cite{buca2019}, not only in the presence of decoherence-free subspaces \cite{lidar1998,knill2000,lidar2003,blume2008}, but also as a consequence of nonequilibrium transitions toward many-body dynamical phases \cite{iemini2018,iemini2021}. Much of the interest in these asymptotic non-stationary states is due to the discovery of time-crystals \cite{wilczek2012,shapere2012}. In simple terms, such time-crystal constitutes a non-stationary nonequilibrium phase of matter, in which the long-time dynamics does not reflect a time-translation symmetry of its generator, see e.g.~Refs.~ \cite{wilczek2012,shapere2012,li2012,else2016,khemani2016,choi2017,zhang2017,sacha2017,lazarides2017,gong2018,gambetta2019,zhu2019,dogra2019,buca2019b,chiacchio2019,yao2020,hurtado2020,nicolaou2021,pizzi2021,kessler2021,iemini2018,buca2019,carollo2020,iemini2021,buonaiuto2021,piccitto2021}.

\begin{figure*}[t]
\centering
\includegraphics[width=0.98\textwidth]{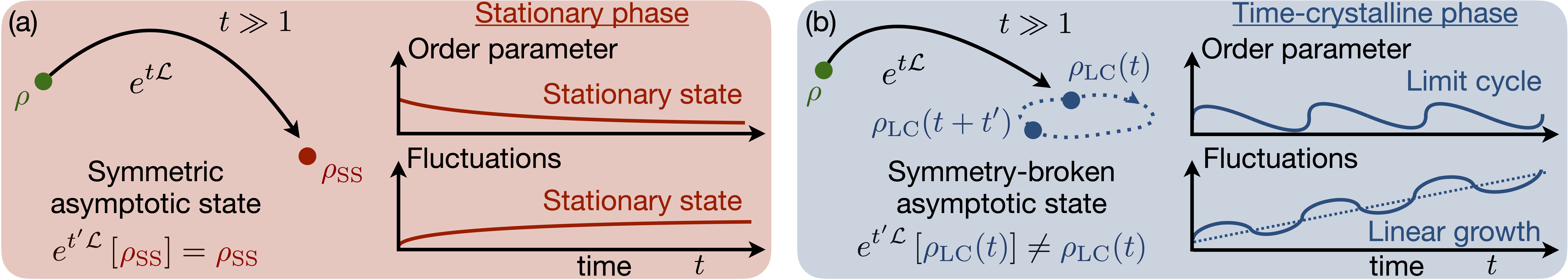}
\caption{{\bf Stationary vs time-crystalline phase.} (a) A Markovian open quantum dynamics  ---implemented by a time-independent generator $\mathcal{L}$--- typically brings the system towards an asymptotic time-invariant state, $\rho_\mathrm{SS}$. Such a stationary state reflects the symmetry of $\mathcal{L}$, since time-translations do not change its properties, $e^{t'\mathcal{L}}[\rho_\mathrm{SS}]=\rho_\mathrm{SS}$. In this regime, system observables, as well as fluctuations, converge to a stationary value, at least far from critical points. (b) In a (continuous) time-crystalline phase, Markovian open quantum systems approach a limit cycle with time-dependent state $\rho_\mathrm{LC}(t)$. Here, the symmetry of the generator is broken given that continuous time-translations modify the position of the quantum state in the limit cycle, i.e.~$e^{t'\mathcal{L}}[\rho_\mathrm{LC}(t)]=\rho_\mathrm{LC}(t+t')\neq \rho_\mathrm{LC}(t)$. In this case, an appropriate  observable ---order parameter--- can witness the persistent oscillations of the asymptotic quantum state. In this paper, we show that in the presence of long-range (collective) dissipative effects, continuous time-crystalline phases feature a critical growth of fluctuations of the order parameter. }
\label{Fig1}
\end{figure*}

Specifically, in the context of Markovian open quantum systems the quantum state obeys the master equation $\dot{\rho}(t)=\mathcal{L}[\rho(t)]$ \cite{lindblad1976,gorini1976,breuer2002,gardiner2004,alicki2007}, with time-independent dynamical generator $\mathcal{L}$. The formal solution,  $\rho(t)=e^{t\mathcal{L}}[\rho(0)]$, introduces the {\it time-translation operator} $e^{t\mathcal{L}}$ which propagates the system for a time $t$. 
Since $\mathcal{L}$ is time independent, one has $[e^{t\mathcal{L}},\mathcal{L}]=0$, showing that time-translation is a continuous ``symmetry" of the generator. 
In these settings, the state $\rho(t)$ is expected to approach a time-independent stationary state $\rho_{\rm SS}$ [see sketch in Fig.~\ref{Fig1}(a)]. 
Such a state reflects the symmetry of the generator, since $e^{t'\mathcal{L}}[\rho_{\rm SS}]=\rho_{\rm SS}$, and may be regarded as a symmetric ``ground state" of $\mathcal{L}$. However, Markovian open quantum systems can also feature non-stationary asymptotic behavior \cite{buca2019}. For large times, the state may approach a limit cycle, i.e.~an asymptotic regime characterized by sustained (usually periodic) oscillations, $\rho(t)\rightarrow \rho_{\rm LC}(t)$ [see Fig.~\ref{Fig1}(b)]. The ``absolute position" within the limit cycle is thus relevant, and further time-translations generically modify the quantum state, $e^{t'\mathcal{L}}[\rho_{\rm LC}(t)]=\rho_{\rm LC}(t'+t)$. The continuous time-translation symmetry of the generator is thus broken and the system forms a crystalline structure in time.

Paradigmatic models featuring a {\it dissipative} continuous time-crystal phase transition are the so-called {\it boundary time-crystals} \cite{iemini2018,carollo2020,iemini2021,buonaiuto2021,piccitto2021}. 
For these models, it has been numerically demonstrated that average (mean-field) operators ---acting as order parameter--- show asymptotic limit-cycle dynamics  [cf.~Fig.~\ref{Fig1}(b)] \cite{iemini2018,iemini2021}. However, an analytic understanding of their behavior, as well as of the behavior of fluctuations and correlations, is still missing, leaving the characterization of these phases incomplete. Here, we provide an exact solution of a boundary time-crystal phase transition. We derive analytical expressions for the order parameter as well as for the dynamics of quantum fluctuations. The latter becomes asymptotically Markovian in the stationary phase, while it remains asymptotically non-Markovian in the time-crystalline phase, as witnessed by the persistent time-dependence of the dynamical generator \cite{chruscinski2010}. Our results show that continuous time-crystals are critical many-body phases, displaying an algebraic growth of fluctuations with time [cf.~Fig.~\ref{Fig1}(b)], typically observed at critical points of second-order phase transitions \cite{onsager1944,fisher1998,hinrichsen2000,gallavotti2013}. \vspace{10pt}

\noindent {\bf The model.---} We consider the original boundary time-crystal model introduced in Ref.~\cite{iemini2018}, which consists of an ensemble of $N$ quantum spin-$1/2$ systems. As a basis for the single-particle algebra, we consider the set $\{v_\alpha\}_{\alpha=0}^{3}$ of (rescaled) Pauli matrices, $v_\alpha=\sigma_{\alpha}/\sqrt{2}$, with $\sigma_0$ proportional to the identity. With the notation $v_\alpha^{(k)}$ we denote the operator $v_\alpha$ acting on the $k$th particle. The model is defined in terms of collective operators $V_\alpha=\sum_{k=1}^N v_\alpha^{(k)}$ and its dynamics is governed a Lindblad dynamical generator \cite{lindblad1976} of the form \cite{iemini2018}
\begin{equation}
\mathcal{L}^*[o]=i[H,o]+\!\!\sum_{\alpha,\beta=1}^{3} \!\!\frac{C_{\alpha\beta}}{N}\! \left(V_\alpha o V_\beta-\frac{1}{2}\left\{o,V_\alpha V_\beta \right\}\right) ,
 \label{gen-obs}   
\end{equation}
which yields the evolution of an observable $o$, through the Heisenberg equation  $\dot{o}(t)=\mathcal{L}^*[o(t)]$. We note that the map $\mathcal{L}^*$ is the dual of $\mathcal{L}$, introduced above. The Hamiltonian $H$ solely consists of single-particle terms and is written as $H=(\omega/\sqrt{2})V_1$, with coherent rate (Rabi frequency) $\omega$ a real number. The second term in Eq.~\eqref{gen-obs} accounts for dissipative contributions, which are parametrized by the matrix $C$. This matrix must be positive semi-definite and can be decomposed into a symmetric part, $A=A^T$, and an anti-symmetric one, $B=-B^T$, as $C=A+iB$. For the model considered, we have that $A_{11}=A_{22}=\gamma$ and $B_{21}=-B_{12}=\gamma$, with all other elements being zero. This yields long-range dissipation which can be formulated in terms of collective jump operators of the form $J=\sqrt{\gamma}(V_1-iV_2)$. The scaling $1/N$ appearing in Eq.~\eqref{gen-obs} ensures the existence of a well-defined thermodynamic ($N\to\infty$) limit \cite{benatti2018}.

Before proceeding, we note that other boundary time-crystal models have been recently proposed \cite{iemini2021,piccitto2021}. These systems belong to a class of models subject to a dissipative collective dynamics generalizing Eq.~\eqref{gen-obs} to higher dimensional single-particle algebras. This class of open quantum systems has been thoroughly investigated in Ref.~\cite{benatti2018}, to which we refer for a  mathematical discussion of the methodology employed here. 
\vspace{10pt}

\noindent {\bf Order parameter.---} To detect the emergence of a time-crystalline phase, we need to identify an appropriate order parameter, see Fig.~\ref{Fig1}(b). The usual choice falls on ``mean-field" operators $m_\alpha^N=V_\alpha/N$ \cite{iemini2018,iemini2021}, accounting for collective macroscopic properties of the system. For initial clustering states, i.e.~states with short-range correlations, the time-evolved operators $m_\alpha^N(t)$ converge, in the large $N$ limit, to multiples of the identity $m_\alpha^N(t)\to m_\alpha(t)=\lim_{N\to\infty}\langle m_\alpha^N(t)\rangle$ \cite{lanford1969,benatti2018}, where  $\langle \cdot \rangle =\Tr(\rho \, \cdot)$ is the quantum expectation. As such, mean-field operators provide a collective dynamical description of the many-body system in terms of classical variables. 

Under the dynamics generated by the map $\mathcal{L}^*$ in Eq.~\eqref{gen-obs}, the evolution of mean-field operators is implemented, in the thermodynamic limit, by a system of nonlinear differential equations \cite{benatti2018,carollo2021}. These equations are
\begin{equation}
\begin{split}
 \dot{m}_{1}(t)&=\gamma\sqrt{2}m_1(t)m_3(t)\, ,\\
  \dot{m}_{2}(t)&=\gamma\sqrt{2}m_2(t)m_3(t)-\omega m_3(t)\, ,\\
   \dot{m}_{3}(t)&=\omega m_2(t) -\gamma \sqrt{2}\left[m_1^2(t)+m_2^2(t)\right]\, ,
 \label{mean-field-eqs}
 \end{split}
\end{equation}
and need to be solved with initial conditions $m_\alpha(0)=\bar{m}_\alpha$. From now on, we set $\gamma=1$ which amounts to measuring time in units of $1/\gamma$ and the energy scale $\omega$ in units of $\gamma$. The norm of the vector $m(t)=[m_1(t),m_2(t),m_3(t)]$ is a conserved quantity. The above equations also feature another conserved quantity \cite{iemini2018} and we consider here initial states with $\bar{m}_1=0$, which implies $m_1(t)=0$ $\forall t$, and $\|m(t)\|^2=1/2$. 

To solve the system in Eq.~\eqref{mean-field-eqs}, we make the ansatz
\begin{equation}
\begin{split}
    m_2(t)&=\cos [f(t)]\bar{m}_2+\sin[f(t)]\bar{m}_3\, ,\\
    m_3(t)&=\cos [f(t)]\bar{m}_3-\sin[f(t)]\bar{m}_2\, .
\end{split}
    \label{solution}
\end{equation}
and substitute this in Eq.~\eqref{mean-field-eqs} to obtain a self-consistent differential equation for $f(t)$. This function completely determines the long-time behavior of the model (see Supplemental Material \cite{SM} for details). 

\begin{figure*}[t]
\centering
\includegraphics[width=0.93\textwidth]{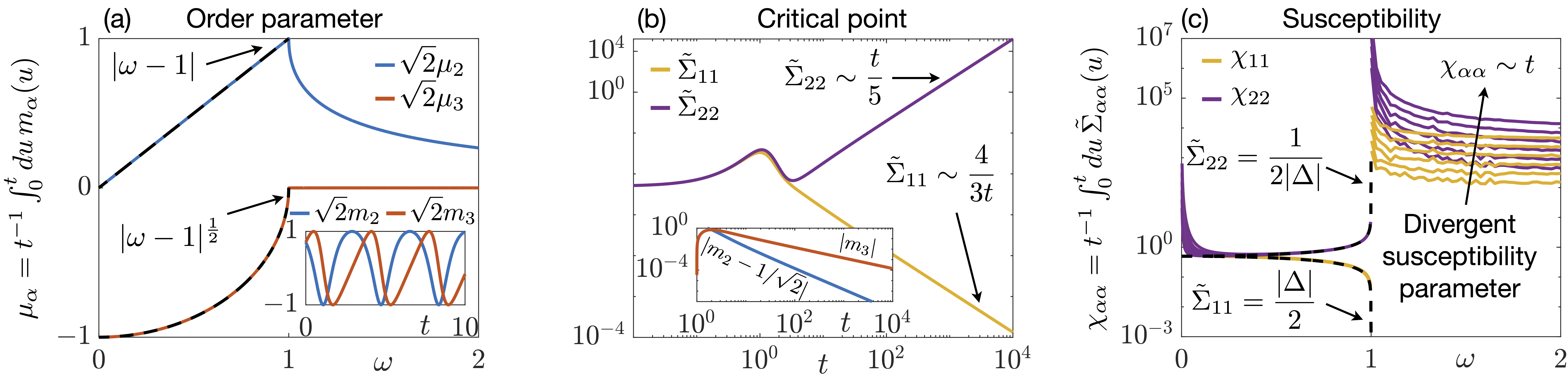}
\caption{{\bf Order parameter and susceptibility.} (a) Time-averaged order parameter $\mu_\alpha$, for $t\to\infty$, from an initial state with $m_2=m_3=1/2$. When the coherent rate $\omega<1$, $\mu_\alpha$  converges to the stationary value $m_\alpha(\infty)$ (dashed lines). Close to criticality, the stationary order parameter components show power-law behavior in $|\omega-1|$ with different exponents. For $\omega>1$, we observe limit-cycle oscillations (shown in the inset for $\omega=2$). The oscillations of $m_3$ average to zero, as shown by $\mu_3$, while $\mu_2$ assumes an $\omega$-dependent value. (b) At the critical point, $\omega=1$, the order parameter components show different power-law decays [see Eq.~\eqref{mf-crit} and inset in log-log scale]. The fluctuation $\tilde{\Sigma}_{11}$ algebraically tends to zero (large spin-squeezing) while the fluctuation $\tilde{\Sigma}_{22}$ diverges with a linear behavior in time, demonstrating a critical building up of (classical) correlations. (c) Linear-log plot of the susceptibility parameters $\chi_{\alpha\alpha}$ as a function of $\omega$, for different times $t=2^i \cdot 500$, with $i=0,1,2,\dots 5$. For $\omega<1$, $\chi_{\alpha\alpha}$ converges to $\tilde{\Sigma}_{\alpha\alpha}(\infty)$ (dashed lines). In the time-crystalline phase, the susceptibility increases linearly with time.}
\label{Fig2}
\end{figure*}

For $\omega<1$, the function $f(t)$ converges to a fixed value $f(\infty):=\lim_{t\to\infty}f(t)$. As such, we have that the limit $m_\alpha(\infty):=\lim_{t\to\infty}m_{\alpha}(t)$ exists for all $\alpha$, ensuring that the system has approached a stationary state [cf.~Fig.~\ref{Fig2}(a)]. In particular, the stationary value of the order parameter components is given by $m_2(\infty)=\omega/\sqrt{2}$ and $m_3(\infty)=-|\Delta|/\sqrt{2}$, with $\Delta=\sqrt{\omega^2-1}$, as reported in Fig.~\ref{Fig2}(a). At the critical coherent rate (Rabi frequency) $\omega=1$, we observe the algebraic behavior (for $t\gg 1$)
\begin{equation}
m_2(t)-m_2(\infty) \sim -\frac{\sqrt{2}}{ t^2}\, \quad \mbox{ and } \quad m_3(t)\sim-\frac{\sqrt{2}}{ t} \, .
    \label{mf-crit}
\end{equation}
On the other hand, for $\omega>1$, the function $f(t)$ does not converge and $m_{2/3}(t)$ feature persistent oscillations witnessing the emergence of a time-crystalline phase [cf.~Fig.~\ref{Fig1}(b)]. In order to retain a phenomenology similar to phase transitions among stationary phases we consider the time-averaged order parameter  $\mu_{\alpha}(t)=t^{-1}\int_0^t du \, m_\alpha(u)$. 
In Fig.~\ref{Fig2}(a), we show that the asymptotic behavior of $\mu_2$ and $\mu_3$ suggests, that this boundary time-crystal phase transition effectively is a continuous second-order transition. 
\vspace{10pt}

\noindent {\bf Quantum fluctuations and correlations.---} The boundary time-crystal phase transition manifests through persistent oscillations of the order parameter. Such a phenomenology is well-known from simple non-linear dynamical systems, such as classical anharmonic oscillators \cite{vanderpol1926}. However, the boundary time-crystal is a genuine many-body phenomenon with surprisingly non-trivial properties. This becomes evident when focusing on quantum fluctuations and on many-body correlations. Their dynamical behavior is not simply inherited from the non-linear (mean-field) system of equations \eqref{mean-field-eqs}.

To study them we introduce the fluctuation operators \cite{goderis1989a,goderis1989b,goderis1990,verbeure2010,benatti2016,benatti2017,benatti2018}
\begin{equation}
    F_\alpha^N=N^{-1/2}\left(V_\alpha -\langle V_\alpha\rangle\right)\, , \, \mbox{ for } \, \alpha=1,2,3\,, 
    \label{fluc-oper}
\end{equation}
which, by definition, account for fluctuations of the operators $V_\alpha$ around their average. Contrary to mean-field operators, they provide a collective description of the many-body system which retains a quantum character, i.e., the limiting operators $F_\alpha:=\lim_{N\to\infty} F_\alpha^N$ are indeed bosonic operators and not classical variables. Their commutation relations are $[F_\alpha,F_\beta]=i \Omega_{\alpha\beta}$ with the matrix  $\Omega_{\alpha\beta}=\sum_{\gamma} \sqrt{2}\epsilon_{\alpha\beta\delta}m_\delta$ and $\epsilon_{\alpha\beta\delta}$ being the fully anti-symmetric tensor. 
For a rigorous discussion about the convergence of quantum fluctuations to bosonic operators, see e.g.~Refs.~\cite{verbeure2010,benatti2017,benatti2018}. 

The rationale for considering fluctuation operators is two-fold: first, they account for two-body correlations, and, second, they quantify fluctuations of the order parameter. In practice, fluctuations provide the susceptibility parameter for the $m_\alpha$, as becomes evident, for example, from the variance of $F_\alpha$:
\begin{equation}
\langle F_\alpha^2\rangle =\lim_{N\to\infty}\frac{1}{N} \sum_{k,h=1}^N \left(\langle v_\alpha^{(k)}v_\alpha^{(h)}\rangle -\langle v_\alpha^{(k)}\rangle \langle v_\alpha^{(h)}\rangle\right) \, .
\label{var-F}
\end{equation}

The time-evolution of quantum fluctuations under the generator in Eq.~\eqref{gen-obs} has been rigorously derived in Ref.~\cite{benatti2018}. Fluctuations undergo a Gaussian dissipative dynamics \cite{heinosaari2010} and their full information is contained in the covariance matrix ${\Sigma}_{\alpha\beta}=\left\langle \left\{F_\alpha ,F_\beta\right\}\right\rangle/2$ \cite{benatti2018}. The precise structure of the dynamics is complicated by the fact that commutation relations between fluctuation operators, as specified by the matrix $\Omega$, are in principle time dependent. This gives rise to an emergent hybrid quantum-classical dynamical system, formed by quantum fluctuations and (classical) mean-field operators \cite{benatti2018}. As we discuss here, this problem can be simplified by looking at the quantum fluctuations $\tilde{F}_\alpha$ defined in the frame rotating with the mean-field operators (see e.g.~Refs.~\cite{pappalardi2018,lerose2020,lerose2020b} for a similar approach in closed systems). 

For completeness, we now briefly sketch the main technical steps for the derivation of the dynamical generator for quantum fluctuations [cf.~Eqs.~\eqref{gen-bos}-\eqref{fluc-dyn-BTC}]. As transparent from Eq.~\eqref{solution}, the time-evolution of mean-field operators can be written through a matrix $R(t)$ as  $m(t)=R(t)m(0)$ (see also the general discussion in Ref.~\cite{benatti2018}). For the model considered, we have  
\begin{equation}
    R(t)=\begin{pmatrix}
    1&0&0\\
    0&\cos[f(t)]&\sin[f(t)]\\
    0&-\sin[f(t)]&\cos[f(t)]
    \end{pmatrix}\, .
    \label{R-matrix}
\end{equation}
The time-evolved covariance matrix $\tilde{\Sigma}(t)$ in the frame rotating with the mean-field operators, can be obtained by subtracting from ${\Sigma}(t)$ the evolution implemented by the unitary matrix $R(t)$, as $\tilde{\Sigma}(t)=R^T(t) \Sigma(t) R(t)$. This covariance matrix obeys the differential equation \cite{SM}
\begin{equation}
\dot{\tilde{\Sigma}}(t)=Q(t)\tilde{\Sigma}(t)+ \tilde{\Sigma}(t)Q^T(t)+\Omega A(t) \Omega^T\, 
    \label{cov-rot-frame}
\end{equation}
with $Q(t)=\Omega B(t)$, where we have used the relation $K(t)=R^{T}(t)KR(t)$, for $K=A,B$. Note, that the matrix $\Omega$ is time independent and fixed to its initial value since, in this frame, mean-field operators do not evolve. 
Remarkably, for the time-evolution of fluctuations in Eq.~\eqref{cov-rot-frame}, we can derive a proper bosonic dynamical generator. We find that the dynamics of any operator $O$ is implemented by a propagator $\Lambda_t$, $O(t)=\Lambda_t\left[O\right]$, obeying the equation  $\dot{\Lambda}_t=\Lambda_t\circ \mathcal{W}_t^*$, with  $\mathcal{W}_t^*$ being the time-dependent Lindblad generator 
\begin{equation}
\begin{split}
    \mathcal{W}^*_t[O]=\sum_{\alpha,\beta=1}^{3}C_{\alpha\beta}(t)\left[ \tilde{F}_\alpha O \tilde{F}_\beta -\frac{1}{2}\left\{\tilde{F}_\alpha \tilde{F}_\beta, O\right\}\right]\, ,
    \end{split}
    \label{gen-bos}
\end{equation}
where $C(t)=A(t)+iB(t)$. This result is rather general and can be adapted to the collective spin models in \cite{benatti2018}.

For the subsequent analysis we use as initial state the one with all spins aligned with the positive direction of $\sigma_3$. Given that we look at the system from the frame rotating with the mean-field variables, we have that $[\tilde{F}_1,\tilde{F}_2]=i$ for all times, so that we can proceed with the identification $\tilde{F}_1=x$ and $\tilde{F}_2=p$, where $x$ and $p$ behave as position and momentum operators, respectively. We also find that $\tilde{F}_3=0$ since $\langle \tilde{F}_3\rangle =0$ and $\langle \tilde{F}_3^2\rangle=0$. Exploiting the result reported in Eq.~\eqref{cov-rot-frame}, we find that the dynamics of quantum fluctuations is governed by the time-dependent generator 
\begin{equation}
\mathcal{W}^*_t[O]= J_t^\dagger O J_t-\frac{1}{2}\left\{J^\dagger_t J_t,O\right\}\, ,
    \label{fluc-dyn-BTC}
\end{equation}
with the ``jump operator" $J_t=x-i\cos[f(t)] p$. This generator clearly shows that fate of quantum fluctuations is also strongly linked with the asymptotic behavior of the time-dependent function $f(t)$. 

For $\omega<1$ the function $f(t)$ rapidly converges to a stationary value so that the generator $\mathcal{W}^*_t$ becomes asymptotically Markovian and we can easily compute the asymptotic state of quantum fluctuations, which is a squeezed vacuum state \cite{SM}. This implies that this phase features spin squeezing \cite{kitagawa1993,ma2011,gross2012} with a squeezing parameter $\xi=|\Delta|=|\sqrt{\omega^2-1}|$ (see also Ref.~\cite{buonaiuto2021}). 

At criticality, $\omega=1$, we find the following behavior for fluctuations (elements of the covariance matrix in the rotated frame)  
\begin{equation}
\tilde{\Sigma}_{11}(t)\sim \frac{4}{3t} \, ,\qquad \tilde{\Sigma}_{22}(t)\sim \frac{ t}{5}\,.
\label{fluc-crit}
\end{equation}
The element  $\tilde{\Sigma}_{11}$ tends to zero [see also Fig.~\ref{Fig2}(b)] and thus shows that, at criticality, the spin-squeezing parameter $\xi$ tends to zero. This is witnessing the presence of strong quantum correlations in the spin ensemble. On the other hand, the element $\tilde{\Sigma}_{22}$ diverges linearly with time, as shown in Fig.~\ref{Fig2}(b). This growth of fluctuations is associated with the divergence of correlations in the ensemble, which usually occurs in second-order phase transitions. 

For $\omega>1$, the generator in Eq.~\eqref{fluc-dyn-BTC} remains time-dependent, signalling that the dynamics of fluctuations is effectively non-Markovian \cite{chruscinski2010}. In this regime, fluctuations show interesting critical behavior. For a stationary nonequilibrium phase transition, one would expect fluctuations to remain bounded away from the critical point, i.e. when $\omega\neq1$. However, we find that the whole time-crystalline regime is characterized by divergent fluctuations. This is associated with the fact that the sustained oscillations of the mean-field operators determine, through the (dissipative) driving term in  $\mathcal{W}^*_t$, an effective ``diffusion" of fluctuations. This behavior is markedly different from an exponential ``heating" of fluctuations which would instead be related to an instability of the mean-field behavior \cite{benatti2016}. Specifically, we find 
\begin{equation}
\begin{split}
\tilde{\Sigma}_{11}(t)&\sim \frac{\left(\Delta^4+\Delta^2\right) t}{2\left(\Delta^2+2\cos^2\left[\frac{\Delta( t+k)}{2}\right]-\Delta\sin\left[\Delta ( t+k)\right]\right)^2}\, ,\\
\tilde{\Sigma}_{22}(t)&\sim \frac{\left(2\Delta^4+5\Delta^2+3\right) t}{2\left(\Delta^2+2\cos^2\left[\frac{\Delta( t+k)}{2}\right]-\Delta\sin\left[\Delta ( t+k)\right]\right)^2}\, ,
\end{split}
    \label{fluct-time-cryst}
\end{equation}
where $k=2\tan^{-1}(1/\Delta)/\Delta$ \cite{SM}. The above quantities, which are well-defined for $\Delta\neq0$, show an overall linear growth of fluctuations  with time. To see this, 
we define the time-averaged susceptibility parameters
$\chi_{\alpha\alpha}(t)= t^{-1}\int_0^t du \, \hat{\Sigma}_{\alpha\alpha}(u)$, which is plotted in Fig.~\ref{Fig2}(c). In the stationary regime, this converges to the stationary value $\tilde{\Sigma}_{\alpha\alpha}(\infty)$. For $\omega>1$, however, these susceptibility parameters diverge with time, indicating that the boundary time-crystal phase is characterized by a critical (unbounded) build-up of correlations which usually solely occurs at a phase transition point. \vspace{10pt}

\noindent {\bf Discussion.---} We have provided an exact solution for the paradigmatic boundary time-crystal model introduced in Ref.~\cite{iemini2018}. In the thermodynamic limit, the dynamical behavior of the order parameter is captured by a set of nonlinear differential equations [see Eq.~\eqref{mean-field-eqs}], which, in certain regimes, can feature persistent oscillations. Our analytical solution shows that a boundary time-crystal is indeed an intricate many-body phase whose physics is much richer than that of a (classical) non-linear system. Remarkably, we have shown that the breaking of the time-translation symmetry becomes manifest in the quantum fluctuations, which evolve under a non-Markovian dynamics (in the sense of Ref.~\cite{chruscinski2010}) in the time-crystalline phase. In contrast, in the stationary phase their dynamics is asymptotically Markovian. Moreover, the analysis of the long-time behavior of quantum fluctuations reveals that the boundary time-crystalline phase is characterized by a diverging power-law growth (with dynamical exponent $1$) of correlations. It thus appears that the entire time-crystal phase is in fact critical.
\\
\begin{acknowledgements}
\noindent \textbf{Acknowledgments.} The research leading to these results has received funding from the European Union’s H2020 research and innovation programme [Grant Agreement No. 800942 (ErBeStA)]. We acknowledge support from the ``Wissenschaftler R\"{u}ckkehrprogramm GSO/CZS" of the Carl-Zeiss-Stiftung and the German Scholars Organization e.V., and the Deutsche Forschungsgemeinschaft through Grants No. 435696605 and 449905436. 
\end{acknowledgements}
\let\oldaddcontentsline\addcontentsline
\renewcommand{\addcontentsline}[3]{}
\bibliography{references}

\begin{thebibliography}{61}%
\makeatletter
\providecommand \@ifxundefined [1]{%
 \@ifx{#1\undefined}
}%
\providecommand \@ifnum [1]{%
 \ifnum #1\expandafter \@firstoftwo
 \else \expandafter \@secondoftwo
 \fi
}%
\providecommand \@ifx [1]{%
 \ifx #1\expandafter \@firstoftwo
 \else \expandafter \@secondoftwo
 \fi
}%
\providecommand \natexlab [1]{#1}%
\providecommand \enquote  [1]{``#1''}%
\providecommand \bibnamefont  [1]{#1}%
\providecommand \bibfnamefont [1]{#1}%
\providecommand \citenamefont [1]{#1}%
\providecommand \href@noop [0]{\@secondoftwo}%
\providecommand \href [0]{\begingroup \@sanitize@url \@href}%
\providecommand \@href[1]{\@@startlink{#1}\@@href}%
\providecommand \@@href[1]{\endgroup#1\@@endlink}%
\providecommand \@sanitize@url [0]{\catcode `\\12\catcode `\$12\catcode
  `\&12\catcode `\#12\catcode `\^12\catcode `\_12\catcode `\%12\relax}%
\providecommand \@@startlink[1]{}%
\providecommand \@@endlink[0]{}%
\providecommand \url  [0]{\begingroup\@sanitize@url \@url }%
\providecommand \@url [1]{\endgroup\@href {#1}{\urlprefix }}%
\providecommand \urlprefix  [0]{URL }%
\providecommand \Eprint [0]{\href }%
\providecommand \doibase [0]{https://doi.org/}%
\providecommand \selectlanguage [0]{\@gobble}%
\providecommand \bibinfo  [0]{\@secondoftwo}%
\providecommand \bibfield  [0]{\@secondoftwo}%
\providecommand \translation [1]{[#1]}%
\providecommand \BibitemOpen [0]{}%
\providecommand \bibitemStop [0]{}%
\providecommand \bibitemNoStop [0]{.\EOS\space}%
\providecommand \EOS [0]{\spacefactor3000\relax}%
\providecommand \BibitemShut  [1]{\csname bibitem#1\endcsname}%
\let\auto@bib@innerbib\@empty
\bibitem [{\citenamefont {Rigol}\ \emph {et~al.}(2008)\citenamefont {Rigol},
  \citenamefont {Dunjko},\ and\ \citenamefont {Olshanii}}]{rigol2008}%
  \BibitemOpen
  \bibfield  {author} {\bibinfo {author} {\bibfnamefont {M.}~\bibnamefont
  {Rigol}}, \bibinfo {author} {\bibfnamefont {V.}~\bibnamefont {Dunjko}},\ and\
  \bibinfo {author} {\bibfnamefont {M.}~\bibnamefont {Olshanii}},\ }\bibfield
  {title} {\bibinfo {title} {Thermalization and its mechanism for generic
  isolated quantum systems},\ }\href {https://doi.org/10.1038/nature06838}
  {\bibfield  {journal} {\bibinfo  {journal} {Nature}\ }\textbf {\bibinfo
  {volume} {452}},\ \bibinfo {pages} {854} (\bibinfo {year}
  {2008})}\BibitemShut {NoStop}%
\bibitem [{\citenamefont {Polkovnikov}\ \emph {et~al.}(2011)\citenamefont
  {Polkovnikov}, \citenamefont {Sengupta}, \citenamefont {Silva},\ and\
  \citenamefont {Vengalattore}}]{polkovnikov2011}%
  \BibitemOpen
  \bibfield  {author} {\bibinfo {author} {\bibfnamefont {A.}~\bibnamefont
  {Polkovnikov}}, \bibinfo {author} {\bibfnamefont {K.}~\bibnamefont
  {Sengupta}}, \bibinfo {author} {\bibfnamefont {A.}~\bibnamefont {Silva}},\
  and\ \bibinfo {author} {\bibfnamefont {M.}~\bibnamefont {Vengalattore}},\
  }\bibfield  {title} {\bibinfo {title} {Colloquium: Nonequilibrium dynamics of
  closed interacting quantum systems},\ }\href
  {https://doi.org/10.1103/RevModPhys.83.863} {\bibfield  {journal} {\bibinfo
  {journal} {Rev. Mod. Phys.}\ }\textbf {\bibinfo {volume} {83}},\ \bibinfo
  {pages} {863} (\bibinfo {year} {2011})}\BibitemShut {NoStop}%
\bibitem [{\citenamefont {D'Alessio}\ \emph {et~al.}(2016)\citenamefont
  {D'Alessio}, \citenamefont {Kafri}, \citenamefont {Polkovnikov},\ and\
  \citenamefont {Rigol}}]{dalessio2016}%
  \BibitemOpen
  \bibfield  {author} {\bibinfo {author} {\bibfnamefont {L.}~\bibnamefont
  {D'Alessio}}, \bibinfo {author} {\bibfnamefont {Y.}~\bibnamefont {Kafri}},
  \bibinfo {author} {\bibfnamefont {A.}~\bibnamefont {Polkovnikov}},\ and\
  \bibinfo {author} {\bibfnamefont {M.}~\bibnamefont {Rigol}},\ }\bibfield
  {title} {\bibinfo {title} {From quantum chaos and eigenstate thermalization
  to statistical mechanics and thermodynamics},\ }\href
  {https://doi.org/10.1080/00018732.2016.1198134} {\bibfield  {journal}
  {\bibinfo  {journal} {Advances in Physics}\ }\textbf {\bibinfo {volume}
  {65}},\ \bibinfo {pages} {239} (\bibinfo {year} {2016})}\BibitemShut
  {NoStop}%
\bibitem [{\citenamefont {Bu{\v c}a}\ \emph {et~al.}(2019)\citenamefont {Bu{\v
  c}a}, \citenamefont {Tindall},\ and\ \citenamefont {Jaksch}}]{buca2019}%
  \BibitemOpen
  \bibfield  {author} {\bibinfo {author} {\bibfnamefont {B.}~\bibnamefont
  {Bu{\v c}a}}, \bibinfo {author} {\bibfnamefont {J.}~\bibnamefont {Tindall}},\
  and\ \bibinfo {author} {\bibfnamefont {D.}~\bibnamefont {Jaksch}},\
  }\bibfield  {title} {\bibinfo {title} {Non-stationary coherent quantum
  many-body dynamics through dissipation},\ }\href
  {https://doi.org/10.1038/s41467-019-09757-y} {\bibfield  {journal} {\bibinfo
  {journal} {Nat. Commun.}\ }\textbf {\bibinfo {volume} {10}},\ \bibinfo
  {pages} {1730} (\bibinfo {year} {2019})}\BibitemShut {NoStop}%
\bibitem [{\citenamefont {Lidar}\ \emph {et~al.}(1998)\citenamefont {Lidar},
  \citenamefont {Chuang},\ and\ \citenamefont {Whaley}}]{lidar1998}%
  \BibitemOpen
  \bibfield  {author} {\bibinfo {author} {\bibfnamefont {D.~A.}\ \bibnamefont
  {Lidar}}, \bibinfo {author} {\bibfnamefont {I.~L.}\ \bibnamefont {Chuang}},\
  and\ \bibinfo {author} {\bibfnamefont {K.~B.}\ \bibnamefont {Whaley}},\
  }\bibfield  {title} {\bibinfo {title} {Decoherence-free subspaces for quantum
  computation},\ }\href {https://doi.org/10.1103/PhysRevLett.81.2594}
  {\bibfield  {journal} {\bibinfo  {journal} {Phys. Rev. Lett.}\ }\textbf
  {\bibinfo {volume} {81}},\ \bibinfo {pages} {2594} (\bibinfo {year}
  {1998})}\BibitemShut {NoStop}%
\bibitem [{\citenamefont {Knill}\ \emph {et~al.}(2000)\citenamefont {Knill},
  \citenamefont {Laflamme},\ and\ \citenamefont {Viola}}]{knill2000}%
  \BibitemOpen
  \bibfield  {author} {\bibinfo {author} {\bibfnamefont {E.}~\bibnamefont
  {Knill}}, \bibinfo {author} {\bibfnamefont {R.}~\bibnamefont {Laflamme}},\
  and\ \bibinfo {author} {\bibfnamefont {L.}~\bibnamefont {Viola}},\ }\bibfield
   {title} {\bibinfo {title} {Theory of quantum error correction for general
  noise},\ }\href {https://doi.org/10.1103/PhysRevLett.84.2525} {\bibfield
  {journal} {\bibinfo  {journal} {Phys. Rev. Lett.}\ }\textbf {\bibinfo
  {volume} {84}},\ \bibinfo {pages} {2525} (\bibinfo {year}
  {2000})}\BibitemShut {NoStop}%
\bibitem [{\citenamefont {Lidar}\ and\ \citenamefont
  {Birgitta~Whaley}(2003)}]{lidar2003}%
  \BibitemOpen
  \bibfield  {author} {\bibinfo {author} {\bibfnamefont {D.~A.}\ \bibnamefont
  {Lidar}}\ and\ \bibinfo {author} {\bibfnamefont {K.}~\bibnamefont
  {Birgitta~Whaley}},\ }\bibinfo {title} {Decoherence-free subspaces and
  subsystems},\ in\ \href {https://doi.org/10.1007/3-540-44874-8_5} {\emph
  {\bibinfo {booktitle} {Irreversible Quantum Dynamics}}},\ \bibinfo {editor}
  {edited by\ \bibinfo {editor} {\bibfnamefont {F.}~\bibnamefont {Benatti}}\
  and\ \bibinfo {editor} {\bibfnamefont {R.}~\bibnamefont {Floreanini}}}\
  (\bibinfo  {publisher} {Springer Berlin Heidelberg},\ \bibinfo {address}
  {Berlin, Heidelberg},\ \bibinfo {year} {2003})\ pp.\ \bibinfo {pages}
  {83--120}\BibitemShut {NoStop}%
\bibitem [{\citenamefont {Blume-Kohout}\ \emph {et~al.}(2008)\citenamefont
  {Blume-Kohout}, \citenamefont {Ng}, \citenamefont {Poulin},\ and\
  \citenamefont {Viola}}]{blume2008}%
  \BibitemOpen
  \bibfield  {author} {\bibinfo {author} {\bibfnamefont {R.}~\bibnamefont
  {Blume-Kohout}}, \bibinfo {author} {\bibfnamefont {H.~K.}\ \bibnamefont
  {Ng}}, \bibinfo {author} {\bibfnamefont {D.}~\bibnamefont {Poulin}},\ and\
  \bibinfo {author} {\bibfnamefont {L.}~\bibnamefont {Viola}},\ }\bibfield
  {title} {\bibinfo {title} {Characterizing the structure of preserved
  information in quantum processes},\ }\href
  {https://doi.org/10.1103/PhysRevLett.100.030501} {\bibfield  {journal}
  {\bibinfo  {journal} {Phys. Rev. Lett.}\ }\textbf {\bibinfo {volume} {100}},\
  \bibinfo {pages} {030501} (\bibinfo {year} {2008})}\BibitemShut {NoStop}%
\bibitem [{\citenamefont {Iemini}\ \emph {et~al.}(2018)\citenamefont {Iemini},
  \citenamefont {Russomanno}, \citenamefont {Keeling}, \citenamefont
  {Schir\`o}, \citenamefont {Dalmonte},\ and\ \citenamefont
  {Fazio}}]{iemini2018}%
  \BibitemOpen
  \bibfield  {author} {\bibinfo {author} {\bibfnamefont {F.}~\bibnamefont
  {Iemini}}, \bibinfo {author} {\bibfnamefont {A.}~\bibnamefont {Russomanno}},
  \bibinfo {author} {\bibfnamefont {J.}~\bibnamefont {Keeling}}, \bibinfo
  {author} {\bibfnamefont {M.}~\bibnamefont {Schir\`o}}, \bibinfo {author}
  {\bibfnamefont {M.}~\bibnamefont {Dalmonte}},\ and\ \bibinfo {author}
  {\bibfnamefont {R.}~\bibnamefont {Fazio}},\ }\bibfield  {title} {\bibinfo
  {title} {Boundary time crystals},\ }\href
  {https://doi.org/10.1103/PhysRevLett.121.035301} {\bibfield  {journal}
  {\bibinfo  {journal} {Phys. Rev. Lett.}\ }\textbf {\bibinfo {volume} {121}},\
  \bibinfo {pages} {035301} (\bibinfo {year} {2018})}\BibitemShut {NoStop}%
\bibitem [{\citenamefont {Prazeres}\ \emph {et~al.}(2021)\citenamefont
  {Prazeres}, \citenamefont {Souza},\ and\ \citenamefont
  {Iemini}}]{iemini2021}%
  \BibitemOpen
  \bibfield  {author} {\bibinfo {author} {\bibfnamefont {L.~F.~d.}\
  \bibnamefont {Prazeres}}, \bibinfo {author} {\bibfnamefont {L.~d.~S.}\
  \bibnamefont {Souza}},\ and\ \bibinfo {author} {\bibfnamefont
  {F.}~\bibnamefont {Iemini}},\ }\bibfield  {title} {\bibinfo {title} {Boundary
  time crystals in collective $d$-level systems},\ }\href
  {https://doi.org/10.1103/PhysRevB.103.184308} {\bibfield  {journal} {\bibinfo
   {journal} {Phys. Rev. B}\ }\textbf {\bibinfo {volume} {103}},\ \bibinfo
  {pages} {184308} (\bibinfo {year} {2021})}\BibitemShut {NoStop}%
\bibitem [{\citenamefont {Wilczek}(2012)}]{wilczek2012}%
  \BibitemOpen
  \bibfield  {author} {\bibinfo {author} {\bibfnamefont {F.}~\bibnamefont
  {Wilczek}},\ }\bibfield  {title} {\bibinfo {title} {Quantum time crystals},\
  }\href {https://doi.org/10.1103/PhysRevLett.109.160401} {\bibfield  {journal}
  {\bibinfo  {journal} {Phys. Rev. Lett.}\ }\textbf {\bibinfo {volume} {109}},\
  \bibinfo {pages} {160401} (\bibinfo {year} {2012})}\BibitemShut {NoStop}%
\bibitem [{\citenamefont {Shapere}\ and\ \citenamefont
  {Wilczek}(2012)}]{shapere2012}%
  \BibitemOpen
  \bibfield  {author} {\bibinfo {author} {\bibfnamefont {A.}~\bibnamefont
  {Shapere}}\ and\ \bibinfo {author} {\bibfnamefont {F.}~\bibnamefont
  {Wilczek}},\ }\bibfield  {title} {\bibinfo {title} {Classical time
  crystals},\ }\href {https://doi.org/10.1103/PhysRevLett.109.160402}
  {\bibfield  {journal} {\bibinfo  {journal} {Phys. Rev. Lett.}\ }\textbf
  {\bibinfo {volume} {109}},\ \bibinfo {pages} {160402} (\bibinfo {year}
  {2012})}\BibitemShut {NoStop}%
\bibitem [{\citenamefont {Li}\ \emph {et~al.}(2012)\citenamefont {Li},
  \citenamefont {Gong}, \citenamefont {Yin}, \citenamefont {Quan},
  \citenamefont {Yin}, \citenamefont {Zhang}, \citenamefont {Duan},\ and\
  \citenamefont {Zhang}}]{li2012}%
  \BibitemOpen
  \bibfield  {author} {\bibinfo {author} {\bibfnamefont {T.}~\bibnamefont
  {Li}}, \bibinfo {author} {\bibfnamefont {Z.-X.}\ \bibnamefont {Gong}},
  \bibinfo {author} {\bibfnamefont {Z.-Q.}\ \bibnamefont {Yin}}, \bibinfo
  {author} {\bibfnamefont {H.~T.}\ \bibnamefont {Quan}}, \bibinfo {author}
  {\bibfnamefont {X.}~\bibnamefont {Yin}}, \bibinfo {author} {\bibfnamefont
  {P.}~\bibnamefont {Zhang}}, \bibinfo {author} {\bibfnamefont {L.-M.}\
  \bibnamefont {Duan}},\ and\ \bibinfo {author} {\bibfnamefont
  {X.}~\bibnamefont {Zhang}},\ }\bibfield  {title} {\bibinfo {title}
  {Space-time crystals of trapped ions},\ }\href
  {https://doi.org/10.1103/PhysRevLett.109.163001} {\bibfield  {journal}
  {\bibinfo  {journal} {Phys. Rev. Lett.}\ }\textbf {\bibinfo {volume} {109}},\
  \bibinfo {pages} {163001} (\bibinfo {year} {2012})}\BibitemShut {NoStop}%
\bibitem [{\citenamefont {Else}\ \emph {et~al.}(2016)\citenamefont {Else},
  \citenamefont {Bauer},\ and\ \citenamefont {Nayak}}]{else2016}%
  \BibitemOpen
  \bibfield  {author} {\bibinfo {author} {\bibfnamefont {D.~V.}\ \bibnamefont
  {Else}}, \bibinfo {author} {\bibfnamefont {B.}~\bibnamefont {Bauer}},\ and\
  \bibinfo {author} {\bibfnamefont {C.}~\bibnamefont {Nayak}},\ }\bibfield
  {title} {\bibinfo {title} {Floquet time crystals},\ }\href
  {https://doi.org/10.1103/PhysRevLett.117.090402} {\bibfield  {journal}
  {\bibinfo  {journal} {Phys. Rev. Lett.}\ }\textbf {\bibinfo {volume} {117}},\
  \bibinfo {pages} {090402} (\bibinfo {year} {2016})}\BibitemShut {NoStop}%
\bibitem [{\citenamefont {Khemani}\ \emph {et~al.}(2016)\citenamefont
  {Khemani}, \citenamefont {Lazarides}, \citenamefont {Moessner},\ and\
  \citenamefont {Sondhi}}]{khemani2016}%
  \BibitemOpen
  \bibfield  {author} {\bibinfo {author} {\bibfnamefont {V.}~\bibnamefont
  {Khemani}}, \bibinfo {author} {\bibfnamefont {A.}~\bibnamefont {Lazarides}},
  \bibinfo {author} {\bibfnamefont {R.}~\bibnamefont {Moessner}},\ and\
  \bibinfo {author} {\bibfnamefont {S.~L.}\ \bibnamefont {Sondhi}},\ }\bibfield
   {title} {\bibinfo {title} {Phase structure of driven quantum systems},\
  }\href {https://doi.org/10.1103/PhysRevLett.116.250401} {\bibfield  {journal}
  {\bibinfo  {journal} {Phys. Rev. Lett.}\ }\textbf {\bibinfo {volume} {116}},\
  \bibinfo {pages} {250401} (\bibinfo {year} {2016})}\BibitemShut {NoStop}%
\bibitem [{\citenamefont {Choi}\ \emph {et~al.}(2017)\citenamefont {Choi},
  \citenamefont {Choi}, \citenamefont {Landig}, \citenamefont {Kucsko},
  \citenamefont {Zhou}, \citenamefont {Isoya}, \citenamefont {Jelezko},
  \citenamefont {Onoda}, \citenamefont {Sumiya}, \citenamefont {Khemani},
  \citenamefont {von Keyserlingk}, \citenamefont {Yao}, \citenamefont
  {Demler},\ and\ \citenamefont {Lukin}}]{choi2017}%
  \BibitemOpen
  \bibfield  {author} {\bibinfo {author} {\bibfnamefont {S.}~\bibnamefont
  {Choi}}, \bibinfo {author} {\bibfnamefont {J.}~\bibnamefont {Choi}}, \bibinfo
  {author} {\bibfnamefont {R.}~\bibnamefont {Landig}}, \bibinfo {author}
  {\bibfnamefont {G.}~\bibnamefont {Kucsko}}, \bibinfo {author} {\bibfnamefont
  {H.}~\bibnamefont {Zhou}}, \bibinfo {author} {\bibfnamefont {J.}~\bibnamefont
  {Isoya}}, \bibinfo {author} {\bibfnamefont {F.}~\bibnamefont {Jelezko}},
  \bibinfo {author} {\bibfnamefont {S.}~\bibnamefont {Onoda}}, \bibinfo
  {author} {\bibfnamefont {H.}~\bibnamefont {Sumiya}}, \bibinfo {author}
  {\bibfnamefont {V.}~\bibnamefont {Khemani}}, \bibinfo {author} {\bibfnamefont
  {C.}~\bibnamefont {von Keyserlingk}}, \bibinfo {author} {\bibfnamefont
  {N.~Y.}\ \bibnamefont {Yao}}, \bibinfo {author} {\bibfnamefont
  {E.}~\bibnamefont {Demler}},\ and\ \bibinfo {author} {\bibfnamefont {M.~D.}\
  \bibnamefont {Lukin}},\ }\bibfield  {title} {\bibinfo {title} {Observation of
  discrete time-crystalline order in a disordered dipolar many-body system},\
  }\href {https://doi.org/10.1038/nature21426} {\bibfield  {journal} {\bibinfo
  {journal} {Nature}\ }\textbf {\bibinfo {volume} {543}},\ \bibinfo {pages}
  {221} (\bibinfo {year} {2017})}\BibitemShut {NoStop}%
\bibitem [{\citenamefont {Zhang}\ \emph {et~al.}(2017)\citenamefont {Zhang},
  \citenamefont {Hess}, \citenamefont {Kyprianidis}, \citenamefont {Becker},
  \citenamefont {Lee}, \citenamefont {Smith}, \citenamefont {Pagano},
  \citenamefont {Potirniche}, \citenamefont {Potter}, \citenamefont
  {Vishwanath}, \citenamefont {Yao},\ and\ \citenamefont {Monroe}}]{zhang2017}%
  \BibitemOpen
  \bibfield  {author} {\bibinfo {author} {\bibfnamefont {J.}~\bibnamefont
  {Zhang}}, \bibinfo {author} {\bibfnamefont {P.~W.}\ \bibnamefont {Hess}},
  \bibinfo {author} {\bibfnamefont {A.}~\bibnamefont {Kyprianidis}}, \bibinfo
  {author} {\bibfnamefont {P.}~\bibnamefont {Becker}}, \bibinfo {author}
  {\bibfnamefont {A.}~\bibnamefont {Lee}}, \bibinfo {author} {\bibfnamefont
  {J.}~\bibnamefont {Smith}}, \bibinfo {author} {\bibfnamefont
  {G.}~\bibnamefont {Pagano}}, \bibinfo {author} {\bibfnamefont {I.~D.}\
  \bibnamefont {Potirniche}}, \bibinfo {author} {\bibfnamefont {A.~C.}\
  \bibnamefont {Potter}}, \bibinfo {author} {\bibfnamefont {A.}~\bibnamefont
  {Vishwanath}}, \bibinfo {author} {\bibfnamefont {N.~Y.}\ \bibnamefont
  {Yao}},\ and\ \bibinfo {author} {\bibfnamefont {C.}~\bibnamefont {Monroe}},\
  }\bibfield  {title} {\bibinfo {title} {Observation of a discrete time
  crystal},\ }\href {https://doi.org/10.1038/nature21413} {\bibfield  {journal}
  {\bibinfo  {journal} {Nature}\ }\textbf {\bibinfo {volume} {543}},\ \bibinfo
  {pages} {217} (\bibinfo {year} {2017})}\BibitemShut {NoStop}%
\bibitem [{\citenamefont {Sacha}\ and\ \citenamefont
  {Zakrzewski}(2017)}]{sacha2017}%
  \BibitemOpen
  \bibfield  {author} {\bibinfo {author} {\bibfnamefont {K.}~\bibnamefont
  {Sacha}}\ and\ \bibinfo {author} {\bibfnamefont {J.}~\bibnamefont
  {Zakrzewski}},\ }\bibfield  {title} {\bibinfo {title} {Time crystals: a
  review},\ }\href {https://doi.org/10.1088/1361-6633/aa8b38} {\bibfield
  {journal} {\bibinfo  {journal} {Reports on Progress in Physics}\ }\textbf
  {\bibinfo {volume} {81}},\ \bibinfo {pages} {016401} (\bibinfo {year}
  {2017})}\BibitemShut {NoStop}%
\bibitem [{\citenamefont {Lazarides}\ and\ \citenamefont
  {Moessner}(2017)}]{lazarides2017}%
  \BibitemOpen
  \bibfield  {author} {\bibinfo {author} {\bibfnamefont {A.}~\bibnamefont
  {Lazarides}}\ and\ \bibinfo {author} {\bibfnamefont {R.}~\bibnamefont
  {Moessner}},\ }\bibfield  {title} {\bibinfo {title} {Fate of a discrete time
  crystal in an open system},\ }\href
  {https://doi.org/10.1103/PhysRevB.95.195135} {\bibfield  {journal} {\bibinfo
  {journal} {Phys. Rev. B}\ }\textbf {\bibinfo {volume} {95}},\ \bibinfo
  {pages} {195135} (\bibinfo {year} {2017})}\BibitemShut {NoStop}%
\bibitem [{\citenamefont {Gong}\ \emph {et~al.}(2018)\citenamefont {Gong},
  \citenamefont {Hamazaki},\ and\ \citenamefont {Ueda}}]{gong2018}%
  \BibitemOpen
  \bibfield  {author} {\bibinfo {author} {\bibfnamefont {Z.}~\bibnamefont
  {Gong}}, \bibinfo {author} {\bibfnamefont {R.}~\bibnamefont {Hamazaki}},\
  and\ \bibinfo {author} {\bibfnamefont {M.}~\bibnamefont {Ueda}},\ }\bibfield
  {title} {\bibinfo {title} {Discrete time-crystalline order in cavity and
  circuit qed systems},\ }\href
  {https://doi.org/10.1103/PhysRevLett.120.040404} {\bibfield  {journal}
  {\bibinfo  {journal} {Phys. Rev. Lett.}\ }\textbf {\bibinfo {volume} {120}},\
  \bibinfo {pages} {040404} (\bibinfo {year} {2018})}\BibitemShut {NoStop}%
\bibitem [{\citenamefont {Gambetta}\ \emph {et~al.}(2019)\citenamefont
  {Gambetta}, \citenamefont {Carollo}, \citenamefont {Marcuzzi}, \citenamefont
  {Garrahan},\ and\ \citenamefont {Lesanovsky}}]{gambetta2019}%
  \BibitemOpen
  \bibfield  {author} {\bibinfo {author} {\bibfnamefont {F.~M.}\ \bibnamefont
  {Gambetta}}, \bibinfo {author} {\bibfnamefont {F.}~\bibnamefont {Carollo}},
  \bibinfo {author} {\bibfnamefont {M.}~\bibnamefont {Marcuzzi}}, \bibinfo
  {author} {\bibfnamefont {J.~P.}\ \bibnamefont {Garrahan}},\ and\ \bibinfo
  {author} {\bibfnamefont {I.}~\bibnamefont {Lesanovsky}},\ }\bibfield  {title}
  {\bibinfo {title} {Discrete time crystals in the absence of manifest
  symmetries or disorder in open quantum systems},\ }\href
  {https://doi.org/10.1103/PhysRevLett.122.015701} {\bibfield  {journal}
  {\bibinfo  {journal} {Phys. Rev. Lett.}\ }\textbf {\bibinfo {volume} {122}},\
  \bibinfo {pages} {015701} (\bibinfo {year} {2019})}\BibitemShut {NoStop}%
\bibitem [{\citenamefont {Zhu}\ \emph {et~al.}(2019)\citenamefont {Zhu},
  \citenamefont {Marino}, \citenamefont {Yao}, \citenamefont {Lukin},\ and\
  \citenamefont {Demler}}]{zhu2019}%
  \BibitemOpen
  \bibfield  {author} {\bibinfo {author} {\bibfnamefont {B.}~\bibnamefont
  {Zhu}}, \bibinfo {author} {\bibfnamefont {J.}~\bibnamefont {Marino}},
  \bibinfo {author} {\bibfnamefont {N.~Y.}\ \bibnamefont {Yao}}, \bibinfo
  {author} {\bibfnamefont {M.~D.}\ \bibnamefont {Lukin}},\ and\ \bibinfo
  {author} {\bibfnamefont {E.~A.}\ \bibnamefont {Demler}},\ }\bibfield  {title}
  {\bibinfo {title} {Dicke time crystals in driven-dissipative quantum
  many-body systems},\ }\href {https://doi.org/10.1088/1367-2630/ab2afe}
  {\bibfield  {journal} {\bibinfo  {journal} {New Journal of Physics}\ }\textbf
  {\bibinfo {volume} {21}},\ \bibinfo {pages} {073028} (\bibinfo {year}
  {2019})}\BibitemShut {NoStop}%
\bibitem [{\citenamefont {Dogra}\ \emph {et~al.}(2019)\citenamefont {Dogra},
  \citenamefont {Landini}, \citenamefont {Kroeger}, \citenamefont {Hruby},
  \citenamefont {Donner},\ and\ \citenamefont {Esslinger}}]{dogra2019}%
  \BibitemOpen
  \bibfield  {author} {\bibinfo {author} {\bibfnamefont {N.}~\bibnamefont
  {Dogra}}, \bibinfo {author} {\bibfnamefont {M.}~\bibnamefont {Landini}},
  \bibinfo {author} {\bibfnamefont {K.}~\bibnamefont {Kroeger}}, \bibinfo
  {author} {\bibfnamefont {L.}~\bibnamefont {Hruby}}, \bibinfo {author}
  {\bibfnamefont {T.}~\bibnamefont {Donner}},\ and\ \bibinfo {author}
  {\bibfnamefont {T.}~\bibnamefont {Esslinger}},\ }\bibfield  {title} {\bibinfo
  {title} {Dissipation-induced structural instability and chiral dynamics in a
  quantum gas},\ }\href {https://doi.org/10.1126/science.aaw4465} {\bibfield
  {journal} {\bibinfo  {journal} {Science}\ }\textbf {\bibinfo {volume}
  {366}},\ \bibinfo {pages} {1496} (\bibinfo {year} {2019})}\BibitemShut
  {NoStop}%
\bibitem [{\citenamefont {Bu\ifmmode~\check{c}\else \v{c}\fi{}a}\ and\
  \citenamefont {Jaksch}(2019)}]{buca2019b}%
  \BibitemOpen
  \bibfield  {author} {\bibinfo {author} {\bibfnamefont {B.}~\bibnamefont
  {Bu\ifmmode~\check{c}\else \v{c}\fi{}a}}\ and\ \bibinfo {author}
  {\bibfnamefont {D.}~\bibnamefont {Jaksch}},\ }\bibfield  {title} {\bibinfo
  {title} {Dissipation induced nonstationarity in a quantum gas},\ }\href
  {https://doi.org/10.1103/PhysRevLett.123.260401} {\bibfield  {journal}
  {\bibinfo  {journal} {Phys. Rev. Lett.}\ }\textbf {\bibinfo {volume} {123}},\
  \bibinfo {pages} {260401} (\bibinfo {year} {2019})}\BibitemShut {NoStop}%
\bibitem [{\citenamefont {Chiacchio}\ and\ \citenamefont
  {Nunnenkamp}(2019)}]{chiacchio2019}%
  \BibitemOpen
  \bibfield  {author} {\bibinfo {author} {\bibfnamefont {E.~I.~R.}\
  \bibnamefont {Chiacchio}}\ and\ \bibinfo {author} {\bibfnamefont
  {A.}~\bibnamefont {Nunnenkamp}},\ }\bibfield  {title} {\bibinfo {title}
  {Dissipation-induced instabilities of a spinor bose-einstein condensate
  inside an optical cavity},\ }\href
  {https://doi.org/10.1103/PhysRevLett.122.193605} {\bibfield  {journal}
  {\bibinfo  {journal} {Phys. Rev. Lett.}\ }\textbf {\bibinfo {volume} {122}},\
  \bibinfo {pages} {193605} (\bibinfo {year} {2019})}\BibitemShut {NoStop}%
\bibitem [{\citenamefont {Yao}\ \emph {et~al.}(2020)\citenamefont {Yao},
  \citenamefont {Nayak}, \citenamefont {Balents},\ and\ \citenamefont
  {Zaletel}}]{yao2020}%
  \BibitemOpen
  \bibfield  {author} {\bibinfo {author} {\bibfnamefont {N.~Y.}\ \bibnamefont
  {Yao}}, \bibinfo {author} {\bibfnamefont {C.}~\bibnamefont {Nayak}}, \bibinfo
  {author} {\bibfnamefont {L.}~\bibnamefont {Balents}},\ and\ \bibinfo {author}
  {\bibfnamefont {M.~P.}\ \bibnamefont {Zaletel}},\ }\bibfield  {title}
  {\bibinfo {title} {Classical discrete time crystals},\ }\href
  {https://doi.org/10.1038/s41567-019-0782-3} {\bibfield  {journal} {\bibinfo
  {journal} {Nature Physics}\ }\textbf {\bibinfo {volume} {16}},\ \bibinfo
  {pages} {438} (\bibinfo {year} {2020})}\BibitemShut {NoStop}%
\bibitem [{\citenamefont {Hurtado-Guti\'errez}\ \emph
  {et~al.}(2020)\citenamefont {Hurtado-Guti\'errez}, \citenamefont {Carollo},
  \citenamefont {P\'erez-Espigares},\ and\ \citenamefont
  {Hurtado}}]{hurtado2020}%
  \BibitemOpen
  \bibfield  {author} {\bibinfo {author} {\bibfnamefont {R.}~\bibnamefont
  {Hurtado-Guti\'errez}}, \bibinfo {author} {\bibfnamefont {F.}~\bibnamefont
  {Carollo}}, \bibinfo {author} {\bibfnamefont {C.}~\bibnamefont
  {P\'erez-Espigares}},\ and\ \bibinfo {author} {\bibfnamefont {P.~I.}\
  \bibnamefont {Hurtado}},\ }\bibfield  {title} {\bibinfo {title} {Building
  continuous time crystals from rare events},\ }\href
  {https://doi.org/10.1103/PhysRevLett.125.160601} {\bibfield  {journal}
  {\bibinfo  {journal} {Phys. Rev. Lett.}\ }\textbf {\bibinfo {volume} {125}},\
  \bibinfo {pages} {160601} (\bibinfo {year} {2020})}\BibitemShut {NoStop}%
\bibitem [{\citenamefont {Nicolaou}\ and\ \citenamefont
  {Motter}(2021)}]{nicolaou2021}%
  \BibitemOpen
  \bibfield  {author} {\bibinfo {author} {\bibfnamefont {Z.~G.}\ \bibnamefont
  {Nicolaou}}\ and\ \bibinfo {author} {\bibfnamefont {A.~E.}\ \bibnamefont
  {Motter}},\ }\bibfield  {title} {\bibinfo {title} {Anharmonic classical time
  crystals: A coresonance pattern formation mechanism},\ }\href
  {https://doi.org/10.1103/PhysRevResearch.3.023106} {\bibfield  {journal}
  {\bibinfo  {journal} {Phys. Rev. Research}\ }\textbf {\bibinfo {volume}
  {3}},\ \bibinfo {pages} {023106} (\bibinfo {year} {2021})}\BibitemShut
  {NoStop}%
\bibitem [{\citenamefont {Pizzi}\ \emph {et~al.}(2021)\citenamefont {Pizzi},
  \citenamefont {Nunnenkamp},\ and\ \citenamefont {Knolle}}]{pizzi2021}%
  \BibitemOpen
  \bibfield  {author} {\bibinfo {author} {\bibfnamefont {A.}~\bibnamefont
  {Pizzi}}, \bibinfo {author} {\bibfnamefont {A.}~\bibnamefont {Nunnenkamp}},\
  and\ \bibinfo {author} {\bibfnamefont {J.}~\bibnamefont {Knolle}},\
  }\bibfield  {title} {\bibinfo {title} {Bistability and time crystals in
  long-ranged directed percolation},\ }\href
  {https://doi.org/10.1038/s41467-021-21259-4} {\bibfield  {journal} {\bibinfo
  {journal} {Nature Communications}\ }\textbf {\bibinfo {volume} {12}},\
  \bibinfo {pages} {1061} (\bibinfo {year} {2021})}\BibitemShut {NoStop}%
\bibitem [{\citenamefont {Ke\ss{}ler}\ \emph {et~al.}(2021)\citenamefont
  {Ke\ss{}ler}, \citenamefont {Kongkhambut}, \citenamefont {Georges},
  \citenamefont {Mathey}, \citenamefont {Cosme},\ and\ \citenamefont
  {Hemmerich}}]{kessler2021}%
  \BibitemOpen
  \bibfield  {author} {\bibinfo {author} {\bibfnamefont {H.}~\bibnamefont
  {Ke\ss{}ler}}, \bibinfo {author} {\bibfnamefont {P.}~\bibnamefont
  {Kongkhambut}}, \bibinfo {author} {\bibfnamefont {C.}~\bibnamefont
  {Georges}}, \bibinfo {author} {\bibfnamefont {L.}~\bibnamefont {Mathey}},
  \bibinfo {author} {\bibfnamefont {J.~G.}\ \bibnamefont {Cosme}},\ and\
  \bibinfo {author} {\bibfnamefont {A.}~\bibnamefont {Hemmerich}},\ }\bibfield
  {title} {\bibinfo {title} {Observation of a dissipative time crystal},\
  }\href {https://doi.org/10.1103/PhysRevLett.127.043602} {\bibfield  {journal}
  {\bibinfo  {journal} {Phys. Rev. Lett.}\ }\textbf {\bibinfo {volume} {127}},\
  \bibinfo {pages} {043602} (\bibinfo {year} {2021})}\BibitemShut {NoStop}%
\bibitem [{\citenamefont {Carollo}\ \emph {et~al.}(2020)\citenamefont
  {Carollo}, \citenamefont {Brandner},\ and\ \citenamefont
  {Lesanovsky}}]{carollo2020}%
  \BibitemOpen
  \bibfield  {author} {\bibinfo {author} {\bibfnamefont {F.}~\bibnamefont
  {Carollo}}, \bibinfo {author} {\bibfnamefont {K.}~\bibnamefont {Brandner}},\
  and\ \bibinfo {author} {\bibfnamefont {I.}~\bibnamefont {Lesanovsky}},\
  }\bibfield  {title} {\bibinfo {title} {Nonequilibrium many-body quantum
  engine driven by time-translation symmetry breaking},\ }\href
  {https://doi.org/10.1103/PhysRevLett.125.240602} {\bibfield  {journal}
  {\bibinfo  {journal} {Phys. Rev. Lett.}\ }\textbf {\bibinfo {volume} {125}},\
  \bibinfo {pages} {240602} (\bibinfo {year} {2020})}\BibitemShut {NoStop}%
\bibitem [{\citenamefont {Buonaiuto}\ \emph {et~al.}(2021)\citenamefont
  {Buonaiuto}, \citenamefont {Carollo}, \citenamefont {Olmos},\ and\
  \citenamefont {Lesanovsky}}]{buonaiuto2021}%
  \BibitemOpen
  \bibfield  {author} {\bibinfo {author} {\bibfnamefont {G.}~\bibnamefont
  {Buonaiuto}}, \bibinfo {author} {\bibfnamefont {F.}~\bibnamefont {Carollo}},
  \bibinfo {author} {\bibfnamefont {B.}~\bibnamefont {Olmos}},\ and\ \bibinfo
  {author} {\bibfnamefont {I.}~\bibnamefont {Lesanovsky}},\ }\bibfield  {title}
  {\bibinfo {title} {Dynamical phases and quantum correlations in an
  emitter-waveguide system with feedback},\ }\href
  {https://doi.org/10.1103/PhysRevLett.127.133601} {\bibfield  {journal}
  {\bibinfo  {journal} {Phys. Rev. Lett.}\ }\textbf {\bibinfo {volume} {127}},\
  \bibinfo {pages} {133601} (\bibinfo {year} {2021})}\BibitemShut {NoStop}%
\bibitem [{\citenamefont {Piccitto}\ \emph {et~al.}(2021)\citenamefont
  {Piccitto}, \citenamefont {Wauters}, \citenamefont {Nori},\ and\
  \citenamefont {Shammah}}]{piccitto2021}%
  \BibitemOpen
  \bibfield  {author} {\bibinfo {author} {\bibfnamefont {G.}~\bibnamefont
  {Piccitto}}, \bibinfo {author} {\bibfnamefont {M.}~\bibnamefont {Wauters}},
  \bibinfo {author} {\bibfnamefont {F.}~\bibnamefont {Nori}},\ and\ \bibinfo
  {author} {\bibfnamefont {N.}~\bibnamefont {Shammah}},\ }\bibfield  {title}
  {\bibinfo {title} {Symmetries and conserved quantities of boundary time
  crystals in generalized spin models},\ }\href
  {https://doi.org/10.1103/PhysRevB.104.014307} {\bibfield  {journal} {\bibinfo
   {journal} {Phys. Rev. B}\ }\textbf {\bibinfo {volume} {104}},\ \bibinfo
  {pages} {014307} (\bibinfo {year} {2021})}\BibitemShut {NoStop}%
\bibitem [{\citenamefont {Lindblad}(1976)}]{lindblad1976}%
  \BibitemOpen
  \bibfield  {author} {\bibinfo {author} {\bibfnamefont {G.}~\bibnamefont
  {Lindblad}},\ }\bibfield  {title} {\bibinfo {title} {{On the generators of
  quantum dynamical semigroups}},\ }\href {https://doi.org/cmp/1103899849}
  {\bibfield  {journal} {\bibinfo  {journal} {Communications in Mathematical
  Physics}\ }\textbf {\bibinfo {volume} {48}},\ \bibinfo {pages} {119 }
  (\bibinfo {year} {1976})}\BibitemShut {NoStop}%
\bibitem [{\citenamefont {Gorini}\ \emph {et~al.}(1976)\citenamefont {Gorini},
  \citenamefont {Kossakowski},\ and\ \citenamefont {Sudarshan}}]{gorini1976}%
  \BibitemOpen
  \bibfield  {author} {\bibinfo {author} {\bibfnamefont {V.}~\bibnamefont
  {Gorini}}, \bibinfo {author} {\bibfnamefont {A.}~\bibnamefont
  {Kossakowski}},\ and\ \bibinfo {author} {\bibfnamefont {E.~C.~G.}\
  \bibnamefont {Sudarshan}},\ }\bibfield  {title} {\bibinfo {title} {Completely
  positive dynamical semigroups of n-level systems},\ }\href
  {https://doi.org/10.1063/1.522979} {\bibfield  {journal} {\bibinfo  {journal}
  {Journal of Mathematical Physics}\ }\textbf {\bibinfo {volume} {17}},\
  \bibinfo {pages} {821} (\bibinfo {year} {1976})}\BibitemShut {NoStop}%
\bibitem [{\citenamefont {Breuer}\ \emph {et~al.}(2002)\citenamefont {Breuer},
  \citenamefont {Petruccione} \emph {et~al.}}]{breuer2002}%
  \BibitemOpen
  \bibfield  {author} {\bibinfo {author} {\bibfnamefont {H.-P.}\ \bibnamefont
  {Breuer}}, \bibinfo {author} {\bibfnamefont {F.}~\bibnamefont {Petruccione}},
  \emph {et~al.},\ }\href@noop {} {\emph {\bibinfo {title} {The theory of open
  quantum systems}}}\ (\bibinfo  {publisher} {Oxford University Press},\
  \bibinfo {year} {2002})\BibitemShut {NoStop}%
\bibitem [{\citenamefont {Gardiner}\ \emph {et~al.}(2004)\citenamefont
  {Gardiner}, \citenamefont {Zoller},\ and\ \citenamefont
  {Zoller}}]{gardiner2004}%
  \BibitemOpen
  \bibfield  {author} {\bibinfo {author} {\bibfnamefont {C.}~\bibnamefont
  {Gardiner}}, \bibinfo {author} {\bibfnamefont {P.}~\bibnamefont {Zoller}},\
  and\ \bibinfo {author} {\bibfnamefont {P.}~\bibnamefont {Zoller}},\
  }\href@noop {} {\emph {\bibinfo {title} {Quantum noise: a handbook of
  Markovian and non-Markovian quantum stochastic methods with applications to
  quantum optics}}}\ (\bibinfo  {publisher} {Springer Science \& Business
  Media},\ \bibinfo {year} {2004})\BibitemShut {NoStop}%
\bibitem [{\citenamefont {Alicki}\ and\ \citenamefont
  {Lendi}(2007)}]{alicki2007}%
  \BibitemOpen
  \bibfield  {author} {\bibinfo {author} {\bibfnamefont {R.}~\bibnamefont
  {Alicki}}\ and\ \bibinfo {author} {\bibfnamefont {K.}~\bibnamefont {Lendi}},\
  }\href@noop {} {\emph {\bibinfo {title} {Quantum dynamical semigroups and
  applications}}},\ Vol.\ \bibinfo {volume} {717}\ (\bibinfo  {publisher}
  {Springer},\ \bibinfo {year} {2007})\BibitemShut {NoStop}%
\bibitem [{\citenamefont {Chru\ifmmode \acute{s}\else
  \'{s}\fi{}ci\ifmmode~\acute{n}\else \'{n}\fi{}ski}\ and\ \citenamefont
  {Kossakowski}(2010)}]{chruscinski2010}%
  \BibitemOpen
  \bibfield  {author} {\bibinfo {author} {\bibfnamefont {D.}~\bibnamefont
  {Chru\ifmmode \acute{s}\else \'{s}\fi{}ci\ifmmode~\acute{n}\else
  \'{n}\fi{}ski}}\ and\ \bibinfo {author} {\bibfnamefont {A.}~\bibnamefont
  {Kossakowski}},\ }\bibfield  {title} {\bibinfo {title} {Non-markovian quantum
  dynamics: Local versus nonlocal},\ }\href
  {https://doi.org/10.1103/PhysRevLett.104.070406} {\bibfield  {journal}
  {\bibinfo  {journal} {Phys. Rev. Lett.}\ }\textbf {\bibinfo {volume} {104}},\
  \bibinfo {pages} {070406} (\bibinfo {year} {2010})}\BibitemShut {NoStop}%
\bibitem [{\citenamefont {Onsager}(1944)}]{onsager1944}%
  \BibitemOpen
  \bibfield  {author} {\bibinfo {author} {\bibfnamefont {L.}~\bibnamefont
  {Onsager}},\ }\bibfield  {title} {\bibinfo {title} {Crystal statistics. i. a
  two-dimensional model with an order-disorder transition},\ }\href
  {https://doi.org/10.1103/PhysRev.65.117} {\bibfield  {journal} {\bibinfo
  {journal} {Phys. Rev.}\ }\textbf {\bibinfo {volume} {65}},\ \bibinfo {pages}
  {117} (\bibinfo {year} {1944})}\BibitemShut {NoStop}%
\bibitem [{\citenamefont {Fisher}(1998)}]{fisher1998}%
  \BibitemOpen
  \bibfield  {author} {\bibinfo {author} {\bibfnamefont {M.~E.}\ \bibnamefont
  {Fisher}},\ }\bibfield  {title} {\bibinfo {title} {Renormalization group
  theory: Its basis and formulation in statistical physics},\ }\href
  {https://doi.org/10.1103/RevModPhys.70.653} {\bibfield  {journal} {\bibinfo
  {journal} {Rev. Mod. Phys.}\ }\textbf {\bibinfo {volume} {70}},\ \bibinfo
  {pages} {653} (\bibinfo {year} {1998})}\BibitemShut {NoStop}%
\bibitem [{\citenamefont {Hinrichsen}(2000)}]{hinrichsen2000}%
  \BibitemOpen
  \bibfield  {author} {\bibinfo {author} {\bibfnamefont {H.}~\bibnamefont
  {Hinrichsen}},\ }\bibfield  {title} {\bibinfo {title} {Non-equilibrium
  critical phenomena and phase transitions into absorbing states},\ }\href
  {https://doi.org/10.1080/00018730050198152} {\bibfield  {journal} {\bibinfo
  {journal} {Advances in Physics}\ }\textbf {\bibinfo {volume} {49}},\ \bibinfo
  {pages} {815} (\bibinfo {year} {2000})}\BibitemShut {NoStop}%
\bibitem [{\citenamefont {Gallavotti}(2013)}]{gallavotti2013}%
  \BibitemOpen
  \bibfield  {author} {\bibinfo {author} {\bibfnamefont {G.}~\bibnamefont
  {Gallavotti}},\ }\href@noop {} {\emph {\bibinfo {title} {Statistical
  mechanics: A short treatise}}}\ (\bibinfo  {publisher} {Springer Science \&
  Business Media},\ \bibinfo {year} {2013})\BibitemShut {NoStop}%
\bibitem [{\citenamefont {Benatti}\ \emph {et~al.}(2018)\citenamefont
  {Benatti}, \citenamefont {Carollo}, \citenamefont {Floreanini},\ and\
  \citenamefont {Narnhofer}}]{benatti2018}%
  \BibitemOpen
  \bibfield  {author} {\bibinfo {author} {\bibfnamefont {F.}~\bibnamefont
  {Benatti}}, \bibinfo {author} {\bibfnamefont {F.}~\bibnamefont {Carollo}},
  \bibinfo {author} {\bibfnamefont {R.}~\bibnamefont {Floreanini}},\ and\
  \bibinfo {author} {\bibfnamefont {H.}~\bibnamefont {Narnhofer}},\ }\bibfield
  {title} {\bibinfo {title} {Quantum spin chain dissipative mean-field
  dynamics},\ }\href {https://doi.org/10.1088/1751-8121/aacbdb} {\bibfield
  {journal} {\bibinfo  {journal} {Journal of Physics A: Mathematical and
  Theoretical}\ }\textbf {\bibinfo {volume} {51}},\ \bibinfo {pages} {325001}
  (\bibinfo {year} {2018})}\BibitemShut {NoStop}%
\bibitem [{\citenamefont {Lanford}\ and\ \citenamefont
  {Ruelle}(1969)}]{lanford1969}%
  \BibitemOpen
  \bibfield  {author} {\bibinfo {author} {\bibfnamefont {O.~E.}\ \bibnamefont
  {Lanford}}\ and\ \bibinfo {author} {\bibfnamefont {D.}~\bibnamefont
  {Ruelle}},\ }\bibfield  {title} {\bibinfo {title} {Observables at infinity
  and states with short range correlations in statistical mechanics},\ }\href
  {https://doi.org/10.1007/BF01645487} {\bibfield  {journal} {\bibinfo
  {journal} {Communications in Mathematical Physics}\ }\textbf {\bibinfo
  {volume} {13}},\ \bibinfo {pages} {194} (\bibinfo {year} {1969})}\BibitemShut
  {NoStop}%
\bibitem [{\citenamefont {Carollo}\ and\ \citenamefont
  {Lesanovsky}(2021)}]{carollo2021}%
  \BibitemOpen
  \bibfield  {author} {\bibinfo {author} {\bibfnamefont {F.}~\bibnamefont
  {Carollo}}\ and\ \bibinfo {author} {\bibfnamefont {I.}~\bibnamefont
  {Lesanovsky}},\ }\bibfield  {title} {\bibinfo {title} {Exactness of
  mean-field equations for open dicke models with an application to pattern
  retrieval dynamics},\ }\href {https://doi.org/10.1103/PhysRevLett.126.230601}
  {\bibfield  {journal} {\bibinfo  {journal} {Phys. Rev. Lett.}\ }\textbf
  {\bibinfo {volume} {126}},\ \bibinfo {pages} {230601} (\bibinfo {year}
  {2021})}\BibitemShut {NoStop}%
\bibitem [{SM()}]{SM}%
  \BibitemOpen
  \href@noop {} {\bibinfo {title} {{see Supplemental Material for
  details}}}\BibitemShut {NoStop}%
\bibitem [{\citenamefont {van~der Pol}(1926)}]{vanderpol1926}%
  \BibitemOpen
  \bibfield  {author} {\bibinfo {author} {\bibfnamefont {B.}~\bibnamefont
  {van~der Pol}},\ }\bibfield  {title} {\bibinfo {title} {On
  “relaxation-oscillations”},\ }\href
  {https://doi.org/10.1080/14786442608564127} {\bibfield  {journal} {\bibinfo
  {journal} {Phil. Mag.}\ }\textbf {\bibinfo {volume} {2}},\ \bibinfo {pages}
  {978} (\bibinfo {year} {1926})},\ \Eprint
  {https://arxiv.org/abs/https://doi.org/10.1080/14786442608564127}
  {https://doi.org/10.1080/14786442608564127} \BibitemShut {NoStop}%
\bibitem [{\citenamefont {Goderis}\ \emph {et~al.}(1989)\citenamefont
  {Goderis}, \citenamefont {Verbeure},\ and\ \citenamefont
  {Vets}}]{goderis1989a}%
  \BibitemOpen
  \bibfield  {author} {\bibinfo {author} {\bibfnamefont {D.}~\bibnamefont
  {Goderis}}, \bibinfo {author} {\bibfnamefont {A.}~\bibnamefont {Verbeure}},\
  and\ \bibinfo {author} {\bibfnamefont {P.}~\bibnamefont {Vets}},\ }\bibfield
  {title} {\bibinfo {title} {Non-commutative central limits},\ }\href
  {https://doi.org/10.1007/BF00341282} {\bibfield  {journal} {\bibinfo
  {journal} {Prob. Th. Rel. Fields}\ }\textbf {\bibinfo {volume} {82}},\
  \bibinfo {pages} {527} (\bibinfo {year} {1989})}\BibitemShut {NoStop}%
\bibitem [{\citenamefont {Goderis}\ and\ \citenamefont
  {Vets}(1989)}]{goderis1989b}%
  \BibitemOpen
  \bibfield  {author} {\bibinfo {author} {\bibfnamefont {D.}~\bibnamefont
  {Goderis}}\ and\ \bibinfo {author} {\bibfnamefont {P.}~\bibnamefont {Vets}},\
  }\bibfield  {title} {\bibinfo {title} {{Central limit theorem for mixing
  quantum systems and the CCR-algebra of fluctuations}},\ }\href
  {https://doi.org/10.1007/BF01257415} {\bibfield  {journal} {\bibinfo
  {journal} {Commun. Math. Phys.}\ }\textbf {\bibinfo {volume} {122}},\
  \bibinfo {pages} {249} (\bibinfo {year} {1989})}\BibitemShut {NoStop}%
\bibitem [{\citenamefont {Goderis}\ \emph {et~al.}(1990)\citenamefont
  {Goderis}, \citenamefont {Verbeure},\ and\ \citenamefont
  {Vets}}]{goderis1990}%
  \BibitemOpen
  \bibfield  {author} {\bibinfo {author} {\bibfnamefont {D.}~\bibnamefont
  {Goderis}}, \bibinfo {author} {\bibfnamefont {A.}~\bibnamefont {Verbeure}},\
  and\ \bibinfo {author} {\bibfnamefont {P.}~\bibnamefont {Vets}},\ }\bibfield
  {title} {\bibinfo {title} {Dynamics of fluctuations for quantum lattice
  systems},\ }\href {https://doi.org/10.1007/BF02096872} {\bibfield  {journal}
  {\bibinfo  {journal} {Commun. Math. Phys.}\ }\textbf {\bibinfo {volume}
  {128}},\ \bibinfo {pages} {533} (\bibinfo {year} {1990})}\BibitemShut
  {NoStop}%
\bibitem [{\citenamefont {Verbeure}(2010)}]{verbeure2010}%
  \BibitemOpen
  \bibfield  {author} {\bibinfo {author} {\bibfnamefont {A.~F.}\ \bibnamefont
  {Verbeure}},\ }\href@noop {} {\emph {\bibinfo {title} {Many-body boson
  systems: half a century later}}}\ (\bibinfo  {publisher} {Springer},\
  \bibinfo {year} {2010})\BibitemShut {NoStop}%
\bibitem [{\citenamefont {Benatti}\ \emph {et~al.}(2016)\citenamefont
  {Benatti}, \citenamefont {Carollo}, \citenamefont {Floreanini},\ and\
  \citenamefont {Narnhofer}}]{benatti2016}%
  \BibitemOpen
  \bibfield  {author} {\bibinfo {author} {\bibfnamefont {F.}~\bibnamefont
  {Benatti}}, \bibinfo {author} {\bibfnamefont {F.}~\bibnamefont {Carollo}},
  \bibinfo {author} {\bibfnamefont {R.}~\bibnamefont {Floreanini}},\ and\
  \bibinfo {author} {\bibfnamefont {H.}~\bibnamefont {Narnhofer}},\ }\bibfield
  {title} {\bibinfo {title} {Non-markovian mesoscopic dissipative dynamics of
  open quantum spin chains},\ }\href
  {https://doi.org/https://doi.org/10.1016/j.physleta.2015.10.062} {\bibfield
  {journal} {\bibinfo  {journal} {Physics Letters A}\ }\textbf {\bibinfo
  {volume} {380}},\ \bibinfo {pages} {381} (\bibinfo {year}
  {2016})}\BibitemShut {NoStop}%
\bibitem [{\citenamefont {Benatti}\ \emph {et~al.}(2017)\citenamefont
  {Benatti}, \citenamefont {Carollo}, \citenamefont {Floreanini},\ and\
  \citenamefont {Narnhofer}}]{benatti2017}%
  \BibitemOpen
  \bibfield  {author} {\bibinfo {author} {\bibfnamefont {F.}~\bibnamefont
  {Benatti}}, \bibinfo {author} {\bibfnamefont {F.}~\bibnamefont {Carollo}},
  \bibinfo {author} {\bibfnamefont {R.}~\bibnamefont {Floreanini}},\ and\
  \bibinfo {author} {\bibfnamefont {H.}~\bibnamefont {Narnhofer}},\ }\bibfield
  {title} {\bibinfo {title} {Quantum fluctuations in mesoscopic systems},\
  }\href {https://doi.org/10.1088/1751-8121/aa84d2} {\bibfield  {journal}
  {\bibinfo  {journal} {Journal of Physics A: Mathematical and Theoretical}\
  }\textbf {\bibinfo {volume} {50}},\ \bibinfo {pages} {423001} (\bibinfo
  {year} {2017})}\BibitemShut {NoStop}%
\bibitem [{\citenamefont {Heinosaari}\ \emph {et~al.}(2010)\citenamefont
  {Heinosaari}, \citenamefont {Holevo},\ and\ \citenamefont
  {Wolf}}]{heinosaari2010}%
  \BibitemOpen
  \bibfield  {author} {\bibinfo {author} {\bibfnamefont {T.}~\bibnamefont
  {Heinosaari}}, \bibinfo {author} {\bibfnamefont {A.~S.}\ \bibnamefont
  {Holevo}},\ and\ \bibinfo {author} {\bibfnamefont {M.~M.}\ \bibnamefont
  {Wolf}},\ }\bibfield  {title} {\bibinfo {title} {The semigroup structure of
  gaussian channels},\ }\href@noop {} {\bibfield  {journal} {\bibinfo
  {journal} {Quantum Info. Comput.}\ }\textbf {\bibinfo {volume} {10}},\
  \bibinfo {pages} {619–635} (\bibinfo {year} {2010})}\BibitemShut {NoStop}%
\bibitem [{\citenamefont {Pappalardi}\ \emph {et~al.}(2018)\citenamefont
  {Pappalardi}, \citenamefont {Russomanno}, \citenamefont {\ifmmode
  \check{Z}\else \v{Z}\fi{}unkovi\ifmmode~\check{c}\else \v{c}\fi{}},
  \citenamefont {Iemini}, \citenamefont {Silva},\ and\ \citenamefont
  {Fazio}}]{pappalardi2018}%
  \BibitemOpen
  \bibfield  {author} {\bibinfo {author} {\bibfnamefont {S.}~\bibnamefont
  {Pappalardi}}, \bibinfo {author} {\bibfnamefont {A.}~\bibnamefont
  {Russomanno}}, \bibinfo {author} {\bibfnamefont {B.}~\bibnamefont {\ifmmode
  \check{Z}\else \v{Z}\fi{}unkovi\ifmmode~\check{c}\else \v{c}\fi{}}}, \bibinfo
  {author} {\bibfnamefont {F.}~\bibnamefont {Iemini}}, \bibinfo {author}
  {\bibfnamefont {A.}~\bibnamefont {Silva}},\ and\ \bibinfo {author}
  {\bibfnamefont {R.}~\bibnamefont {Fazio}},\ }\bibfield  {title} {\bibinfo
  {title} {Scrambling and entanglement spreading in long-range spin chains},\
  }\href {https://doi.org/10.1103/PhysRevB.98.134303} {\bibfield  {journal}
  {\bibinfo  {journal} {Phys. Rev. B}\ }\textbf {\bibinfo {volume} {98}},\
  \bibinfo {pages} {134303} (\bibinfo {year} {2018})}\BibitemShut {NoStop}%
\bibitem [{\citenamefont {Lerose}\ and\ \citenamefont
  {Pappalardi}(2020{\natexlab{a}})}]{lerose2020}%
  \BibitemOpen
  \bibfield  {author} {\bibinfo {author} {\bibfnamefont {A.}~\bibnamefont
  {Lerose}}\ and\ \bibinfo {author} {\bibfnamefont {S.}~\bibnamefont
  {Pappalardi}},\ }\bibfield  {title} {\bibinfo {title} {Origin of the slow
  growth of entanglement entropy in long-range interacting spin systems},\
  }\href {https://doi.org/10.1103/PhysRevResearch.2.012041} {\bibfield
  {journal} {\bibinfo  {journal} {Phys. Rev. Research}\ }\textbf {\bibinfo
  {volume} {2}},\ \bibinfo {pages} {012041} (\bibinfo {year}
  {2020}{\natexlab{a}})}\BibitemShut {NoStop}%
\bibitem [{\citenamefont {Lerose}\ and\ \citenamefont
  {Pappalardi}(2020{\natexlab{b}})}]{lerose2020b}%
  \BibitemOpen
  \bibfield  {author} {\bibinfo {author} {\bibfnamefont {A.}~\bibnamefont
  {Lerose}}\ and\ \bibinfo {author} {\bibfnamefont {S.}~\bibnamefont
  {Pappalardi}},\ }\bibfield  {title} {\bibinfo {title} {Bridging entanglement
  dynamics and chaos in semiclassical systems},\ }\href
  {https://doi.org/10.1103/PhysRevA.102.032404} {\bibfield  {journal} {\bibinfo
   {journal} {Phys. Rev. A}\ }\textbf {\bibinfo {volume} {102}},\ \bibinfo
  {pages} {032404} (\bibinfo {year} {2020}{\natexlab{b}})}\BibitemShut
  {NoStop}%
\bibitem [{\citenamefont {Kitagawa}\ and\ \citenamefont
  {Ueda}(1993)}]{kitagawa1993}%
  \BibitemOpen
  \bibfield  {author} {\bibinfo {author} {\bibfnamefont {M.}~\bibnamefont
  {Kitagawa}}\ and\ \bibinfo {author} {\bibfnamefont {M.}~\bibnamefont
  {Ueda}},\ }\bibfield  {title} {\bibinfo {title} {Squeezed spin states},\
  }\href {https://doi.org/10.1103/PhysRevA.47.5138} {\bibfield  {journal}
  {\bibinfo  {journal} {Phys. Rev. A}\ }\textbf {\bibinfo {volume} {47}},\
  \bibinfo {pages} {5138} (\bibinfo {year} {1993})}\BibitemShut {NoStop}%
\bibitem [{\citenamefont {Ma}\ \emph {et~al.}(2011)\citenamefont {Ma},
  \citenamefont {Wang}, \citenamefont {Sun},\ and\ \citenamefont
  {Nori}}]{ma2011}%
  \BibitemOpen
  \bibfield  {author} {\bibinfo {author} {\bibfnamefont {J.}~\bibnamefont
  {Ma}}, \bibinfo {author} {\bibfnamefont {X.}~\bibnamefont {Wang}}, \bibinfo
  {author} {\bibfnamefont {C.}~\bibnamefont {Sun}},\ and\ \bibinfo {author}
  {\bibfnamefont {F.}~\bibnamefont {Nori}},\ }\bibfield  {title} {\bibinfo
  {title} {Quantum spin squeezing},\ }\href
  {https://doi.org/https://doi.org/10.1016/j.physrep.2011.08.003} {\bibfield
  {journal} {\bibinfo  {journal} {Physics Reports}\ }\textbf {\bibinfo {volume}
  {509}},\ \bibinfo {pages} {89} (\bibinfo {year} {2011})}\BibitemShut
  {NoStop}%
\bibitem [{\citenamefont {Gross}(2012)}]{gross2012}%
  \BibitemOpen
  \bibfield  {author} {\bibinfo {author} {\bibfnamefont {C.}~\bibnamefont
  {Gross}},\ }\bibfield  {title} {\bibinfo {title} {Spin squeezing,
  entanglement and quantum metrology with bose{\textendash}einstein
  condensates},\ }\href {https://doi.org/10.1088/0953-4075/45/10/103001}
  {\bibfield  {journal} {\bibinfo  {journal} {Journal of Physics B: Atomic,
  Molecular and Optical Physics}\ }\textbf {\bibinfo {volume} {45}},\ \bibinfo
  {pages} {103001} (\bibinfo {year} {2012})}\BibitemShut {NoStop}%
\end{thebibliography}%
\let\addcontentsline\oldaddcontentsline

\newpage

\renewcommand\thesection{S\arabic{section}}
\renewcommand\theequation{S\arabic{equation}}
\renewcommand\thefigure{S\arabic{figure}}
\setcounter{equation}{0}
\setcounter{figure}{0}

\onecolumngrid

\begin{center}
{\Large SUPPLEMENTAL MATERIAL}
\end{center}
\begin{center}
\vspace{0.8cm}
{\Large Exact solution  of a boundary time-crystal phase transition: time-translation symmetry breaking and non-Markovian dynamics of correlations}
\end{center}
\begin{center}
Federico Carollo$^{1}$ and Igor Lesanovsky$^{1,2}$
\end{center}
\begin{center}
$^1${\em Institut f\"ur Theoretische Physik, Universit\"at T\"ubingen,}\\
{\em Auf der Morgenstelle 14, 72076 T\"ubingen, Germany}\\
$^2${\em School of Physics and Astronomy and}\\
{\em Centre for the Mathematics and Theoretical Physics of Quantum Non-Equilibrium Systems,}\\
{\em  University of Nottingham, Nottingham, NG7 2RD, UK}\\

\end{center}


\section{Exact solution of the order-parameter dynamics}
In this section we provide an exact analytical expression for the solution of the mean-field nonlinear equations of motion in the case $m_1=0$. In this case, one has that $m_1(t)=0$, for all $t$, and also that $m_y^2(t)+m_z^2(t)=s^2$ is a conserved quantity. We consider the case of $s=1/\sqrt{2}$. The system of differential equations that needs to be solved is given by
\begin{equation}
\begin{split}
\dot{m}_2(t)&=\sqrt{2}m_2(t)m_3(t)-\omega m_3(t)\\
\dot{m}_3(t)&=\omega m_2(t) -\sqrt{2}m_2^2(t)\, .
\end{split}
    \label{mf-red}
\end{equation}

In order to solve it, we assume the existence of a function $f(t)$ such that 
\begin{equation}
\begin{split}
{m}_2(t)&=\cos [f(t)] \bar{m}_2 +\sin [f(t)] \bar{m}_3 \\
{m}_3(t)&=\cos [f(t)] \bar{m}_3 -\sin [f(t)] \bar{m}_2 \, ,
\end{split}
    \label{ansatz}
\end{equation}
where $\bar{m}_\alpha$ are the initial conditions for the mean-field operators. We then observe that, with the above definition, $\dot{m}_2(t)=\dot{f}(t)m_3(t)$ as well as $\dot{m}_3(t)=-\dot{f}(t)m_2(t)$. Substituting these in Eq.~\eqref{mf-red}, we obtain from both equations the same differential equation for $\dot{f}$, which is the following (we now omit the dependence on $t$ if this is not necessary)
\begin{equation}
    \dot{f}=\sqrt{2}[\cos (f) \bar{m}_2+\sin (f) \bar{m}_3]-\omega\, .
\end{equation}
We thus need to find a proper function $f$ satisfying the above relation. Separating variables we can write 
$$
\frac{df}{\sqrt{2}\bar{m}_2 \cos(f)+\sqrt{2}\bar{m}_3 \sin (f) -\omega}=dt\, ,
$$
which  we integrate  as 
\begin{equation}
I:= \int \frac{df}{\sqrt{2}\bar{m}_2 \cos(f)+\sqrt{2}\bar{m}_3 \sin (f) -\omega} =t+k\, ,
    \label{int-f}
\end{equation}
where $k$ is a constant to be determined. We thus need to solve the integral $I$. To this end, we proceed with the Weierstrass substitution
\begin{equation}
    g=\tan (f/2) \, , \quad \sin(f)=\frac{2g}{1+g^2}\, , \quad \cos (f)=\frac{1-g^2}{1+g^2}\, , \quad df=\frac{2}{1+g^2} dg\, ,
\end{equation}
to obtain
$$
I=-\int dg \frac{2}{ag^2 +bg+c}\, , \qquad \mbox{ with }\quad a=\sqrt{2}\bar{m}_2+\omega \, , \quad b=-2\sqrt{2}\bar{m}_3 \, ,\quad c=\omega -\sqrt{2}\bar{m}_2\, .
$$
The structure of the solution of the above integral depends on the value of the (rescaled) discriminant $\Delta^2=(ac-b^2/4)$. Computing this quantity we find that 
$$
\Delta^2=\omega^2-1\, .
$$
We thus have three possible solutions for $\omega^2<1$ (stationary phase), for $\omega^2=1$ (critical point) and for $\omega^2>1$ (time-crystalline phase). We analyze these separately. 

\subsection{Solution in the stationary phase}
In this case we have $\Delta^2<0$. The solution for the integral $I$ is then given by
$$
I=\frac{2}{|\Delta|}\mathrm{tanh}^{-1}\left(\frac{2ag+b}{2|\Delta|}\right)\, ,
$$
and substituting the expressions of $a,b$ and $g$, we find
$$
I=\frac{2}{|\Delta|}\mathrm{tanh}^{-1}\left(\frac{(\sqrt{2}\bar{m}_2+\omega)\tan (f/2)-\sqrt{2}\bar{m}_3}{|\Delta|}\right)\, .
$$
By setting $I=t+k$ and solving for $f$ we obtain
$$
f=2\tan^{-1}\left(\frac{|\Delta|\tanh \left(\frac{|\Delta|(t+k)}{2}\right)+\sqrt{2}\bar{m}_3}{\sqrt{2}\bar{m}_2+\omega}\right)\, ,
$$
with $k$ fixed by the initial condition $f(0)=0$, which gives 
$$
k=\frac{2}{|\Delta|}\mathrm{tanh}^{-1}\left(-\frac{\sqrt{2}\bar{m}_3}{|\Delta|}\right)\, .
$$

\subsubsection{Asymptotics of the order parameters}
In order to recover the long-time value of the mean-field operators $m_2(\infty)$ and $m_3(\infty)$, we first look at the limit
$$
\lim_{t\to\infty}f(t)=2\tan^{-1}\left[\frac{|\Delta|+\sqrt{2}\bar{m}_3}{\sqrt{2}\bar{m}_2+\omega}\right]\, ,
$$
which converges to a finite stationary value. Using that 
$$
\cos\left[2\tan^{-1}(y)\right]=\frac{1-y^2}{1+y^2}\, , \qquad \mbox{ and } \qquad \sin\left[2\tan^{-1}(y)\right]=\frac{2y}{1+y^2}\, ,
$$
it can be checked that (remember that $\bar{m}_2^2+\bar{m}_3^2=1/2$)
\begin{equation}
\cos\left[f(\infty)\right]=\frac{2(\bar{m}_2^2-\bar{m}_3^2)+2\omega^2-1+2\sqrt{2}(\omega \bar{m}_2-|\Delta|\bar{m}_3)}{2+2\sqrt{2}(\bar{m}_2\omega +\bar{m}_3|\Delta|)}\, , \quad \mbox{ and } \quad \sin\left[f(\infty)\right]=\frac{2(|\Delta|+\sqrt{2}\bar{m}_3)(\sqrt{2}\bar{m}_2+\omega)}{2+2\sqrt{2}(\bar{m}_2\omega +\bar{m}_3|\Delta|)}\, .
\end{equation}
With the above relations, after some lengthy but straightforward calculations, one finds
\begin{equation}
    \begin{split}
    m_2(\infty)&=\cos[f(\infty)]\bar{m}_2+\sin[f(\infty)]\bar{m}_3=\frac{\omega}{\sqrt{2}}\, ,\\
        m_3(\infty)&=\cos[f(\infty)]\bar{m}_3-\sin[f(\infty)]\bar{m}_2=-\frac{|\Delta|}{\sqrt{2}}\, ,
    \end{split}
    \label{asymp-sub}
\end{equation}
which do not depend on the initial values $\bar{m}_2$ and $\bar{m}_3$.

\subsection{Solution for the critical rate}
In this case we have $\Delta=0$, since $\omega=1$. The solution for the integral $I$ is then given by
$$
I=\frac{4}{2ag+b}\, ,
$$
which substituting for $a,b$  and $g$ becomes
$$
I=\frac{4}{2(\sqrt{2}\bar{m}_2+\omega)\tan(f/2)-2\sqrt{2}\bar{m}_3}\, .
$$
Solving for $f$, we then find 
$$
f(t)=2\tan^{-1}\left[\frac{1}{\sqrt{2}\bar{m}_2+\omega}\left(\frac{2}{t+k}+\sqrt{2}\bar{m}_3\right)\right]\, ,
$$
where $k$ is given by 
$$
k=-\frac{\sqrt{2}}{\bar{m}_3}\, .
$$
\subsubsection{Asymptotics of the order parameters}
In this situation we also have that $\lim_{t\to\infty}f(t)$ converges to an asymptotic fixed value. However, in this case the long-time dynamics features an algebraic decay so that we are not really interested in the $t\to\infty$ limit, but we actually want to find the  behavior of the order parameter for large times. 

To this end, we want to expand the function $f(t)$ for large times $t\gg1$ (remember that here rates are in units of $\gamma$ while time is in units of $1/\gamma$). Neglecting the constant $k$, for large $t$ we write 
$$
f(t)\sim 2\tan^{-1}\left[\alpha +\beta/t\right]\, , \quad \mbox{ with }\quad  \alpha=\frac{\sqrt{2}\bar{m}_3}{\sqrt{2}\bar{m}_2+1} \quad \mbox{ and }\quad  \beta=\frac{2}{\sqrt{2}\bar{m}_2+1}
$$
For convenience we now write $f(1/x)$ and expand for small $x$. We thus have
$$
f(1/x)\sim 2\tan^{-1}(\alpha)+\frac{2\beta}{1+\alpha^2}x\, ,
$$
and also 
\begin{equation}
\begin{split}
    \cos [f(1/x)]&\sim \cos[2\tan^{-1}(\alpha)]-\sin[2\tan^{-1}(\alpha)]\frac{2\beta}{1+\alpha^2}x\\
    \sin [f(1/x)]&\sim \sin[2\tan^{-1}(\alpha)]+\cos[2\tan^{-1}(\alpha)]\frac{2\beta}{1+\alpha^2}x\, .
    \end{split}
\end{equation}
Substituting in the expression for $m_3(t)$ as a function of $f$, we arrive at 
\begin{equation}
    m_3(t)\sim \cos[2\tan^{-1}(\alpha)]\bar{m}_3-\sin[2\tan^{-1}(\alpha)]\bar{m}_2-\frac{C}{t}\, ,
    \label{exp-crit}
\end{equation}
where we have introduced the constant 
$$
C=\frac{2\beta}{(1+\alpha^2)^2}\left[2\alpha \bar{m}_3+(1-\alpha^2)\bar{m}_2\right]\, .
$$
The first two terms in Eq.~\eqref{exp-crit} provide the asymptotic value $m_3(\infty)$, which can be evaluated as before [or taking the limit $\Delta\to0$ in Eq.~\eqref{asymp-sub}]. The constant $C$ can be evaluated to $C=\sqrt{2}$. 
Using the second equation in \eqref{mf-red} and the fact that $m_2^2(t)=1/2-m_3^2(t)$ we  find the asymptotic expansion for $m_2(t)$ reported in the main text.

\subsection{Solution in the time-crystalline phase}
In this case we have $\Delta^2>0$. The solution for the integral $I$ is given by 
$$
I=\frac{2}{\Delta}\tan^{-1}\left[\frac{-2ag -b}{2\Delta}\right]\, .
$$
Substituting in the above expression the value of $a,b$ and $g$ we find
$$
I=\frac{2}{\Delta}\tan^{-1}\left[\frac{\sqrt{2}\bar{m}_3-(\sqrt{2}\bar{m}_2+\omega)\tan(f/2)}{\Delta}\right]\, ;
$$
setting $I=t+k$ and inverting the relation to  explicitly find $f$, we obtain 
$$
f(t)=2\tan^{-1}\left[\frac{\sqrt{2}\bar{m}_3-\Delta\tan\left(\frac{\Delta (t+k)}{2}\right)}{\omega +\sqrt{2}\bar{m}_2}\right]\, ,
$$
where $k$ needs to be fixed by insisting that $f(0)=0$, which leads to 
$$
k=\frac{2}{\Delta}\tan^{-1} \left(\frac{\sqrt{2}\bar{m}_3}{\Delta}\right)\, .
$$
\subsubsection{Asymptotics of the order parameters}
In this parameter regime, one can observe that the $\lim_{t\to\infty}f(t)$ is not well-defined since such a function features persistent oscillations. This determines the indefinite oscillations of the order parameters.

\section{Dynamics of fluctuations in the rotating frame}
In this section we demonstrate how to obtain the time-dependent generator for the quantum fluctuations defined in the frame rotating with the mean-field observables.

\subsection{Dynamics of quantum fluctuations in the frame rotating with macroscopic observables}
The dynamics of the quantum fluctuations in a fixed reference frame was derived in Ref.~\cite{benatti2018}. This is accounted for by the dynamics of the covariance matrix $\Sigma(t)$ which obeys the differential equation 
$$
\dot{\Sigma}(t)=W(t)\Sigma(t)+\Sigma(t)W^T(t)+\Omega(t)A\Omega^T(t)\, , \qquad \mbox{ with }\qquad W(t)=\Omega(t)B+D(t)\, ,
$$
where we have 
\begin{equation}
\Omega(t)=\sqrt{2}\begin{pmatrix}
0&m_3(t)&-m_2(t)\\
-m_3(t)&0&m_1(t)\\
m_2(t)&-m_1(t)&0
\end{pmatrix}\, , \qquad \mbox{ and }\qquad  D(t)=\begin{pmatrix}
0&0&0\\
0&0&-\omega +\sqrt{2}m_2(t)\\
0&\omega-\sqrt{2}m_2(t)&0
\end{pmatrix}\, .
    \label{Om-D-t}
\end{equation}
As discussed in the main text,  we can pass from the fixed reference frame to the one rotating with mean-field operators by ``subtracting" to the full evolution of the covariance matrix, the dynamics of mean-field operators. This means that we are interested in the covariance matrix  
$$
\tilde{\Sigma}(t)=R^T(t) \Sigma(t) R(t)\, .
$$
Note that we have $\dot{R}(t)=D(t)R(t)$. To proceed, we first take the time-derivative of the above expression to find the differential equation for the matrix $\tilde{\Sigma}$. We obtain 
$$
\dot{\tilde{\Sigma}}(t)=R^T(t) \left[W(t)-D(t)\right]\tilde{\Sigma}(t)R(t)+R^T(t)\tilde{\Sigma}(t)\left[W^T(t)+D(t)\right]R(t)+R^T(t)\Omega(t) A\Omega^T(t) R(t)\, .
$$
Now, we note that, as discussed in Ref.~\cite{benatti2018}, $R^T(t)\Omega(t)R(t)=\Omega(0)=:\Omega$. Then, we consider the following matrix
$$
R^T(t)\left[W(t)-D(t)\right]R(t)=R^T(t)\left[\Omega(t) B\right]R(t)=\Omega B(t)=:Q(t)\, ,
$$
with $B(t)=R^T(t) B R(t)$. Finally, we also have that $R^T(t)\Omega(t) A\Omega^T(t) R(t)=\Omega A(t)\Omega^T$, with $A(t)=R^T(t) A R(t)$. We can now exploit all the above results, by considering that $R(t)$ is unitary and inserting identities in the form of $1=R(t)R^T(t)$ before the $\tilde{\Sigma}(t)$ in the first term, after $\tilde{\Sigma}(t)$ in the second term, and both before and after the matrix $A$ in the last term. With this we achieve the result
$$
\dot{\tilde{\Sigma}}(t)=Q(t)\tilde{\Sigma}(t) +\tilde{\Sigma}(t) Q^T (t)+\Omega A(t)\Omega^T \, ,
$$
which is the differential equation appearing in Eq.~\eqref{cov-rot-frame} of the main text.

\subsection{Time-dependent (bosonic) generator for quantum fluctuations in the rotating frame}
For the bosonic gaussian dynamics of quantum fluctuations encoded in the covariance matrix $\tilde{\Sigma}(t)$, it is possible to obtain a well-defined time-dependent generator in Lindblad-like form. This can be done by back engineering the proper bosonic dynamical generator which gives rise to the  dynamics of $\tilde{\Sigma}(t)$. As we verify below such a generator assumes the form 
\begin{equation}
\mathcal{W}_t^*[O]=\sum_{\alpha,\beta=1}^{3}C_{\alpha\beta}(t)\left[ \tilde{F}_\alpha O \tilde{F}_\beta -\frac{1}{2}\left\{\tilde{F}_\alpha \tilde{F}_\beta, O\right\}\right]\, ,
\label{bos-gen-SM}
\end{equation}
and acts on any possible operator $O$ constructed from products and sums of the fluctuation operators $\{\tilde{F}_\alpha\}_{\alpha=1}^{3}$. 

Since the generator is time-dependent, the resulting dynamics is not of semi-group type. Therefore, the propagator $\Lambda_t$, implementing the dynamics of an operator $O$ as $O(t)=\Lambda_t[O]$, is not a simple exponential of $\mathcal{W}_t^*$. In fact, we have that
\begin{equation*}
\dot{O}(t)=\Lambda_t\circ \mathcal{W}_t^*[O]\, ,
\end{equation*}
where $\circ$ denotes composition of maps. The above equation shows that $\dot{\Lambda}_t=\Lambda_t\circ \mathcal{W}_t^*$, which needs to be solved with initial condition $\Lambda_0={\rm id}$, where ${\rm id}$ is the identity map. \\

We now demonstrate that the generator in Eq.~\eqref{bos-gen-SM} implements the correct dynamics. First of all, we note that the operator is a quadratic operator ---it must indeed implement a gaussian dynamics--- so that it is sufficient to understand its action on quadratic products of fluctuation operators. Therefore, we now consider $\tilde{F}_\mu \tilde{F}_\nu$.
We start noting  that 
$$
\mathcal{W}_t^*[\tilde{F}_\mu \tilde{F}_\nu]=\mathcal{W}_t^*\left[\tilde{F}_\mu\right]\tilde{F}_\nu+\tilde{F}_\mu\mathcal{W}^*_t\left[\tilde{F}_\nu\right]+\sum_{\alpha,\beta=1}^{d^2-1}C_{\alpha\beta}(t)\left[\tilde{F}_\alpha,\tilde{F}_\mu\right]\left[\tilde{F}_\nu,\tilde{F}_\beta\right]\, .
$$
Then, we have 
$$
\mathcal{W}^*_t\left[\tilde{F}_\mu\right]=-\sum_{\alpha,\beta=1}^{3}B_{\alpha\beta}(t)\, \Omega_{\alpha\mu} \tilde{F}_\beta\, ,
$$
and 
$$
\sum_{\alpha,\beta=1}^{3}C_{\alpha\beta}(t)\left[\tilde{F}_\alpha,\tilde{F}_\mu\right]\left[\tilde{F}_\nu,\tilde{F}_\beta\right]=\left[\Omega \, C(t) \, \Omega^T\right]_{\mu\nu}\, .
$$
Defining the expectation value of the two-point operators as $G_{\mu\nu}= \langle \tilde{F}_\mu \tilde{F}_\nu\rangle$ and using the results above we find that 
$$
\dot{G}(t)=\Omega B(t) G(t) +G(t) B(t)\Omega
+\Omega \, C(t)\, \Omega^T\, .
$$
Since the covariance matrix is given by $\tilde{\Sigma}(t)=\left(G(t)+G^T(t)\right)/2$, the above result implies [remember that $A(t)=(C(t)+C^T(t))/2$]
$$
\dot{\tilde{\Sigma}}(t)=Q(t)\tilde{\Sigma}(t)+\tilde{\Sigma}(t) Q^T(t)+\Omega A(t) \Omega^T\, ,
$$
with $Q(t)=\Omega B(t)$. The above equation is exactly the one for $\tilde{\Sigma}(t)$. This demonstrates that the time-dependent generator $\mathcal{W}_t^*$, as defined in Eq.~\eqref{bos-gen-SM}, implements the dynamics we are after.

The solution for the covariance matrix $\tilde{\Sigma}(t)$ is given by 
\begin{equation}
\tilde{\Sigma}(t)=X_t\Sigma(0) X^T_t+\int_0^t du\, X_t X^{-1}_u \Omega A(t) \Omega^T \left[X_t X^{-1}_u \right]^T\, ,
    \label{sol-bos-CM}
\end{equation}
where $\dot{X}_t=Q(t) X_t$.

\section{Behavior of fluctuations in the different phases}
To study the behavior of quantum fluctuations, we now consider the problem with initial state given by $\bar{m}_1=\bar{m}_2=0$ and $\bar{m}_3=1/\sqrt{2}$. For this situation, the dynamical generator can be reduced to the relevant operators $\tilde{F}_1,\tilde{F}_2$, truncating the sum in Eq.~\eqref{bos-gen-SM}. Here, the matrices assume the form 
\begin{equation}
    A(t)=\begin{pmatrix}1&0\\
    0&\cos^2[f(t)]
    \end{pmatrix}\, ,\qquad B(t)=\begin{pmatrix}0&-\cos[f(t)]\\
    \cos[f(t)]&0
    \end{pmatrix}\, ,\qquad \Omega=\begin{pmatrix}0&1\\
    -1&0
    \end{pmatrix}\, .
\end{equation}

\subsection{Stationary phase}
In the stationary phase, we have that the function $\cos[f(t)]\to  -|\Delta|$ for large times. This is a consequence of the validity of Eq.~\eqref{asymp-sub} for any initial $\bar{m}_2$ and $\bar{m}_3$. The generator in Eq.~\eqref{bos-gen-SM} thus becomes asymptotically Markovian and the stationary state of fluctuations can be obtained by finding the stationary state of the Markovian generator $\mathcal{W}_\infty^*$. 

The stationary state for this dynamics is a squeezed vacuum state characterized by the covariance matrix 
$$
\tilde{\Sigma}(\infty)=\frac{1}{2}\begin{pmatrix}|\Delta|&0\\
0&\frac{1}{|\Delta|}
\end{pmatrix}\, .
$$

\subsection{Critical point}
At the critical value of $\omega=1$, we can compute
$$
\cos[f(t)]=\frac{-4t +4}{2t^2-4t+4}\, ,
$$
for which we find 
$$
X_t=\frac{4}{2t^2-4t+4}\begin{pmatrix}
1&0\\0&1
\end{pmatrix}\, .
$$
Substituting these findings in the solution given by Eq.~\eqref{sol-bos-CM}, we obtain 
$$
\tilde{\Sigma}(t)=\left(\frac{4}{2t^2-4t+4}\right)^2\begin{pmatrix}
\frac{1}{2}&0\\0&\frac{1}{2}
\end{pmatrix}+\begin{pmatrix}
\frac{16/3 t^3-16t^2+16t}{\left(2t^2-4t+4\right)^2}&0\\0&\frac{4/5 t^5-4t^4+32/3t^3-16t^2+16t}{\left(2t^2-4t+4\right)^2}
\end{pmatrix}\, .
$$

\subsection{Time-crystalline phase}
For $\omega>1$, the system is found in the time-crystalline phase. Here, $f(t)$ shows sustained oscillations and for our initial state assumes the form 
$$
f(t)=2\tan^{-1}\left[\frac{1-\Delta \tan\left(\frac{\Delta(t+k)}{2}\right)}{\omega}\right]\, , \qquad \mbox{ with }  \qquad k=\frac{2}{\Delta}\tan^{-1} \left(\frac{1}{\Delta}\right)\, .
$$
We can then express $\cos[f(t)]$ as 
$$
\cos[f(t)]=\frac{\omega^2-\left[1-\Delta \tan\left(\frac{\Delta(t+k)}{2}\right)\right]^2}{\omega^2+\left[1-\Delta \tan\left(\frac{\Delta(t+k)}{2}\right)\right]^2}\, .
$$
Integrating the above quantity and exponentiating we find 
$$
X_t=\frac{\Delta^2 +2\cos^2\left(\frac{\Delta k}{2}\right)-\Delta \sin(\Delta k)}{\Delta^2 +2\cos^2\left[\frac{\Delta (t+k)}{2}\right]-\Delta \sin\left[\Delta (t+k)\right]}\begin{pmatrix}
1&0\\0&1
\end{pmatrix}\, .
$$

Using these, we can express the time-dependent covariance matrix as
\begin{equation*}
\begin{split}
\tilde{\Sigma}(t)&=\left[\frac{\Delta^2 +2\cos^2\left(\frac{\Delta k}{2}\right)-\Delta \sin(\Delta k)}{\Delta^2 +2\cos^2\left[\frac{\Delta (t+k)}{2}\right]-\Delta \sin\left[\Delta (t+k)\right]}\right]^2 \begin{pmatrix}
\frac{1}{2}&0\\0&\frac{1}{2}
\end{pmatrix}+\\
&+\int_0^t du\, \begin{pmatrix}\left[\frac{\Delta^2\cos[\Delta(u+k)]+\Delta \sin[\Delta(u+k)]}{\Delta^2 +2\cos^2\left[\frac{\Delta (t+k)}{2}\right]-\Delta \sin\left[\Delta (t+k)\right]}\right]^2&0\\
0&\left[\frac{\Delta^2 +2\cos^2\left[\frac{\Delta (u+k)}{2}\right]-\Delta \sin\left[\Delta (u+k)\right]}{\Delta^2 +2\cos^2\left[\frac{\Delta (t+k)}{2}\right]-\Delta \sin\left[\Delta (t+k)\right]}\right]^2
\end{pmatrix}\, .
\end{split}
\end{equation*}
We left the numerators in the second term implicitly defined through integrals as the explicit solution, while straightforward, is rather lengthy and not particularly instructive.

\subsubsection{Behavior in the fixed reference frame}
The explicit knowledge of the time-dependent behavior of the covariance matrix $\tilde{\Sigma}(t)$ can be readily exploited to obtain the behavior of fluctuations in the original fixed reference frame. This is simply achieved by reversing the transformation that led to $\tilde{\Sigma}$. We thus have $\Sigma(t)=R(t)\tilde{\Sigma}(t)R^T(t)$, which reads explicitly as
\begin{equation}
\begin{split}
    \Sigma(t)&=\begin{pmatrix}
    1&0&0\\
    0&\cos[f(t)]&\sin[f(t)]\\
    0&-\sin[f(t)]&\cos[f(t)]
    \end{pmatrix} \begin{pmatrix}
    \tilde{\Sigma}_{11}(t)&0&0\\
    0&\tilde{\Sigma}_{22}(t)&0\\
    0&0&0
    \end{pmatrix} \begin{pmatrix}
    1&0&0\\
    0&\cos[f(t)]&-\sin[f(t)]\\
    0&\sin[f(t)]&\cos[f(t)]
    \end{pmatrix}=\\
    &=\begin{pmatrix}
    \tilde{\Sigma}_{11}(t)&0&0\\
    0&\tilde{\Sigma}_{22}(t)\cos^2[f(t)]&-\tilde{\Sigma}_{22}(t)\sin[f(t)]\cos[f(t)]\\
    0&-\tilde{\Sigma}_{22}(t)\sin[f(t)]\cos[f(t)]&\tilde{\Sigma}_{22}(t)\sin^2[f(t)]
    \end{pmatrix}\, .
    \end{split}
\end{equation}

\end{document}